\documentclass{raa}

\usepackage{graphicx,times}
\usepackage{natbib}
\usepackage{ctex}
\usepackage{amssymb,amsmath}
\bibpunct{(}{)}{;}{a}{}{,}

\usepackage[pagebackref=true]{hyperref}

\newcommand{\phrate}{\rm photons~s^{-1}}
\newcommand{\ctrate}{\rm counts~s^{-1}}

\renewcommand{\figurename}{Figure.}
\renewcommand{\tablename}{Table.}
\renewcommand{\abstractname}{Abstract}

\usepackage{tabularx}
\usepackage[version=4]{mhchem}
\usepackage{xcolor}
\usepackage{appendix}
\usepackage{multirow}

\begin{document}

   \title{DIffuse X-ray Explorer (DIXE): Sky Survey Strategy and Collimator Response Demodulation
}

 \volnopage{ {\bf 20XX} Vol.\ {\bf X} No. {\bf XX}, 000--000}
   \setcounter{page}{1}

   \author{Jiejia Liu  
   \inst{1}, Chunyang Jiang\inst{1}, Junjie Mao\inst{1,*}, Rui Huang\inst{1,2}, Ruixuan Tian\inst{1}, Wei Cui\inst{1,*}
      \footnotetext{$*$Corresponding Authors, these authors contributed equally to this work.}
   }

   \institute{Department of Astronomy, Tsinghua University, Beijing 100086,
China; {\it cui@tsinghua.edu.cn}\\
        \and
             Department of Astronomy, University of Michigan, 311 West Hall, 1085 S. University Ave, Ann Arbor, MI 48109-1107, USA\\
   {\small Received 20XX Month Day; accepted 20XX Month Day}
}

\abstract{DIffuse X-ray Explorer (DIXE) is a proposed high-resolution X-ray spectroscopic surveyor aimed at studying large structures of hot gas in the Milky Way. Its payload is designed to have a field of view (FoV) of $10^\circ$ (half-power diameter) and an energy resolution of better than 6 eV, covering an energy range of $0.1-10$ keV. It will be mounted on the China Space Station (CSS) and follow the CSS orbit to conduct the survey with fixed zenith pointing in order to optimize the coverage of key science targets. The payload will avoid the Sun passively via an operable sunshade where a minimum $25^\circ$ angular separation between the pointing axis and the direction of the Sun is required. Two Sun-avoidance strategies are considered: one focusing on minimizing mechanical risk and the other on maximizing exposure time. The one-year exposure maps indicate that DIXE will cover approximately 72.5\% of the sky, with typical exposure times of $26~\rm ks$ and $68~\rm ks$ for the two strategies, respectively. 
Although mechanically collimated, the imaging performance of the payload can be enhanced with a demodulation method based on Markov Chain Monte Carlo sampling using the collimator response. Through simulation, we found that the method could achieve a localization accuracy of $1^\circ$ for point-like sources and a spatial resolution of $3^\circ$ for the extended sources of complex surface brightness distribution, both of which are significantly smaller than the FoV. 
}

   \authorrunning{Jiejia Liu et al. }            
   \titlerunning{DIXE exposure map and collimator response}  
   \maketitle

%
\section{Introduction} \label{sec:intro}

The Milky Way, as our host galaxy, serves as both a crucial template and a natural laboratory for studying hot baryons and the associated physical processes in galaxies. Probing the hot baryons in the Milky Way may offer important insights into accretion and feedback processes and their roles in driving galaxy evolution. The eROSITA all-sky survey \citep{Predehl_eROSIT_2019}
has significantly boosted our understanding of hot gas in the Milky Way, thanks to its improved sensitivity and better energy and spatial resolution compared to the ROSAT all-sky survey \citep{Snowden_1995_ROSAT}. However, the spectral resolving power ($R=E/\Delta E\sim10$) of CCD detectors is still insufficient to resolve emission lines in the X-ray spectrum of hot plasmas. Characteristic emission lines from elements such as C, N, O, Ne, and Fe are crucial diagnostics of plasma temperature, chemical abundances, kinematics, and emission mechanisms.

To fill the observational gap, DIffuse X-ray Explorer (DIXE) was proposed and is to be deployed on the China Space Station (CSS) \citep{Jin_2024_DIXE}. DIXE is to employ a microcalorimeter array as its detector, which looks through a collimated field of view of $10^\circ$ (half-power diameter) and offers an energy resolution of better than 6 eV over an energy range of 0.1-10 keV. Table~\ref{tab:inst_param} summarizes some of the key characteristics.
DIXE is designed to carry out a three-year sky survey with unprecedented spectral resolution, complementing the eROSITA imaging survey. 
The primary science drivers include the Local Hot Bubble (LHB; \citealp{Snowden_1995_ROSAT, Sanders_2001_LHB, Zucker_2020_LHB}), the eROSITA Bubble \citep{Predehl_2020_Nature, Sarkar_2024_eROSITAbubble_review}, the Milky Way halo (MWH; \citealp{Miller_2015_CGM, Henley_2013_XMM_MWH, Ponti_2023_eROSITA_MWH}), as well as other large structures of hot gas (superbubbles and supernova remnants) and solar wind charge exchange (SWCX; \citealp{Cravens_2001_SWCX, Fujimoto_2007_SWCX, Kuntz_2019_SWCX}).
The characterization of X-ray emission from hot gas in the Milky Way with DIXE would also allow accurate foreground modeling in the studies of extragalactic circumgalactic medium (CGM) with future missions, including HUBS \citep{Cui_2020_HUBS} and NewAthena \citep{Cruise_2025_NewAthena}.

It is essential to optimize DIXE's survey strategy by balancing technical risks with scientific returns. Based on the selected strategies, exposure maps can be generated accordingly to evaluate scientific capabilities and provide baselines for forecasting. Although the collimator design defines angular resolution and thus imaging ability,  spatial resolution can be improved to an extent by characterizing the effects of X-ray modulation related to the collimator response \citep{Li_Wu_1994_demodulation,Li_1995_demodulation,2020JHEAp..25...39N}. 
In this paper, we present the optimization of DIXE's survey strategy and apply a demodulation method to enhance its imaging capability.

This paper is organized as follows. In Section~\ref{sec:survey_strategy}, we describe the optimization of survey strategies, including sky coverage and Sun avoidance, and derive one-year exposure maps. 
In Section~\ref{sec:collimator_response}, we introduce a method to demodulate the survey data and retrieve the spatial distribution from the collimator response using Monte Carlo Markov Chain (MCMC). 
We summarize our results in Section~\ref{sec:summary}.

\begin{table}
    \newcolumntype{Y}{>{\centering\arraybackslash}X}
    \renewcommand{\arraystretch}{1.15}
    \centering
    \caption{Summary of the targeted parameters of the DIXE instrument.}
    \begin{tabular}{ll}
    \hline
    \hline
    Parameter & Value \\
    \hline
    Platform &  China Space Station  \\
    Observing mode & Survey \\
    Energy band & 0.1 -- 10 keV \\
    FoV & $10^\circ\times10^\circ$ (collimator) \\
    Detector & Transition Edge Sensor (TES)-based microcalorimeter ($10\times10$ array)\\
    Effective area & $1~\rm cm^2$@$6~\rm keV$ \\
    Grasp & 100 $\rm deg^2~cm^2$@$6~\rm keV$ \\
    Energy resolution &  6 eV@0.6 keV  \\
    Launch time & 2029 (expected) \\
    \hline
    \end{tabular}
    \label{tab:inst_param}
\end{table}










\section{Optimizing survey strategy}
\label{sec:survey_strategy}
\subsection{CSS Orbit and co-moving frame}
\begin{figure}
    \centering
    \includegraphics[width=\hsize]{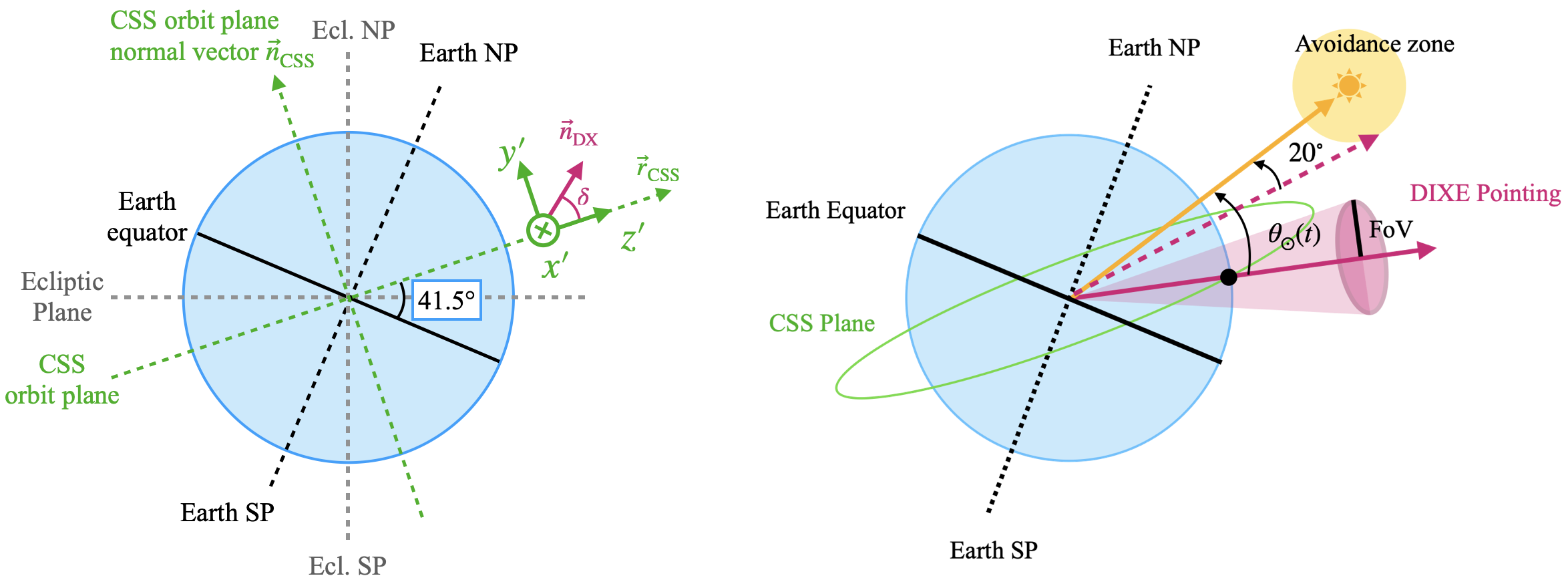}
    \caption{Geometry of the CSS orbit and DIXE pointing direction (not drawn to scale). The Earth equator and its north and south poles are shown. In the \textit{left} panel, the Ecliptic plane and its north and south poles are also shown. The CSS orbital plane is shown in green dashed line, with its normal vector ($\vec n_{\rm CSS}$) and positional vector ($\vec r_{\rm CSS}$) marked. The inclination angle between the CSS plane and the Earth's equator is about $41.5^\circ$. A Cartesian coordinate system ($x'y'z'$) is defined in the CSS co-moving frame. The purple arrow shows the DIXE pointing direction $\vec n_{\rm DX}$, with $\delta$ indicating the angle between it and $z'$-axis. In the \textit{right} panel, the purple cones show the DIXE field of view. The Sun angle, $\theta_\odot(t)$, is defined as the angle between the DIXE pointing vector and the Sun's position vector.
    A Sun avoidance zone is defined and marked in yellow. 
    }
    \label{fig:sunavd}
\end{figure}

The DIXE payload will be mounted on CSS with a fixed pointing direction. The CSS orbits the Earth in a low Earth orbit (LEO), at an altitude between 340 and 420 km \footnote{\url{https://www.n2yo.com/satellite/?s=48274\#google_vignette}}. 
As depicted in Figure~\ref{fig:sunavd} \textit{left} panel, the CSS orbital plane is inclined at an angle of approximately $41.5^\circ$ to the Earth's equator.
The CSS completes an orbit around the Earth approximately every 92 minutes, and its precession period is about 2 months due to the ``westward regression" effect caused by the atmospheric drag \citep{Vallado_2006_spacetrack_report}. 

As shown in Figure~\ref{fig:sunavd} We define the right-handed Cartesian coordinate system in the CSS comoving frame $x'y'z'$ as: 1) The $z'$ axis is along the $\vec r_{\rm CSS}$ direction, i.e., the vector starting from the center of the CSS plane to the position of DIXE; 2) The $y'$ axis is parallel to the normal vector of the CSS plane $\vec n_{\rm CSS}$; 3) The $x'$ axis is parallel to the CSS velocity. The pointing of DIXE $\vec n_{\rm DX}$ is shown by the purple arrow in Figure~\ref{fig:sunavd} with an inclination angle, i.e., the angle with respect to the $z'$-axis, $\delta$. 

To obtain real-time orbital information—including the position and velocity of the CSS—we use the Python package \texttt{sgp4}\footnote{\url{https://pypi.org/project/sgp4/}}, which outputs coordinates in the True Equator Mean Equinox reference frame \citep{Vallado_2006_spacetrack_report}. These are then converted to the Galactic coordinate system using the \texttt{astropy} package \citep{astropy_collab_2013_astropy}. The exposure time is computed and projected onto Galactic coordinates using the \texttt{healpy} package \citep{Healpy_2005}. 
Throughout this work, we base our analyses on the actual CSS orbital information for 2022. 

\subsection{Inclination angle and sky coverage}
\begin{figure}
    \centering
    \includegraphics[width=\hsize]{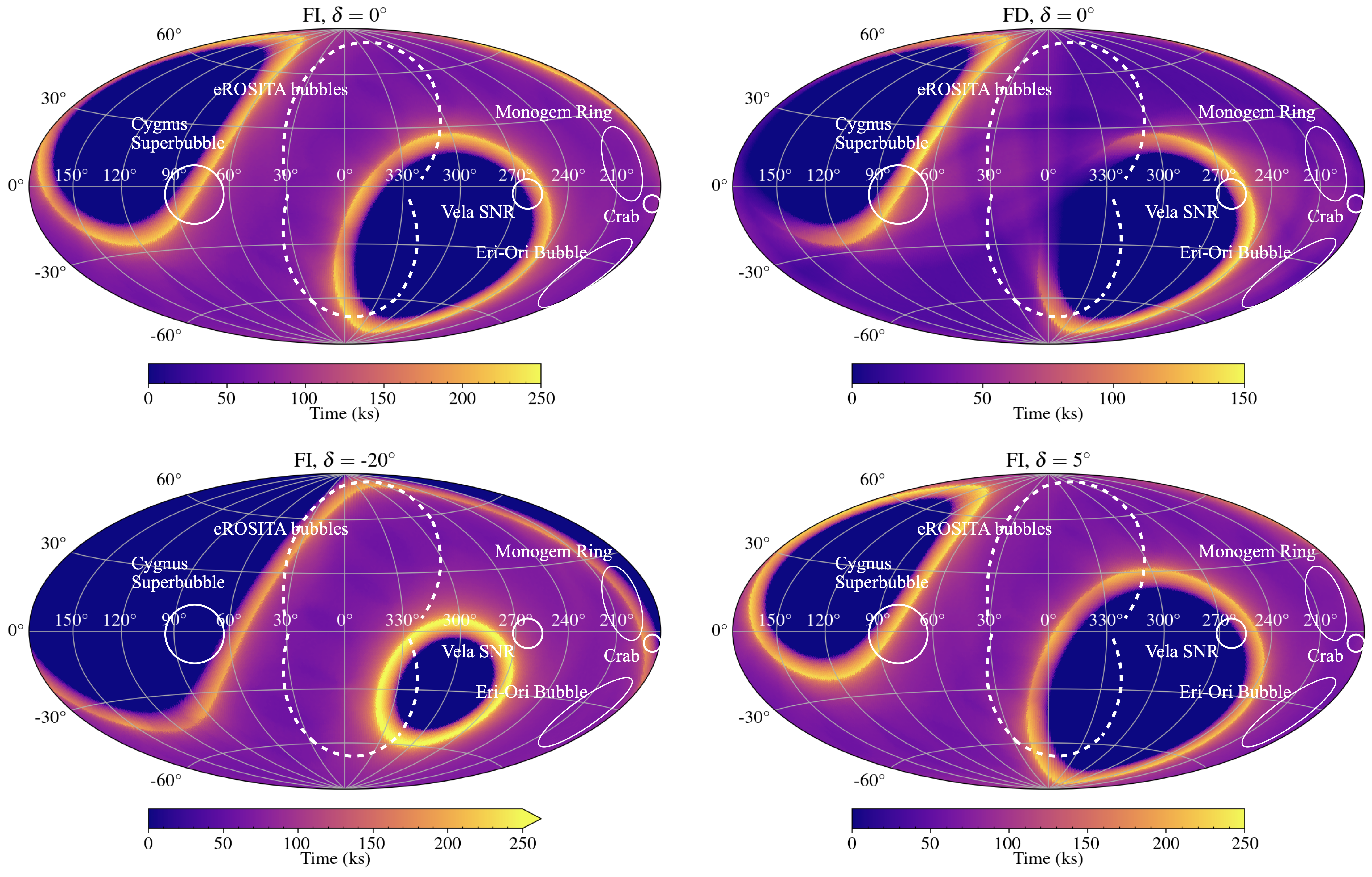}
    \caption{One-year exposure maps under two closure strategies and different inclination angles. The \textit{upper} panels show the exposure map under the forbidden interval (FI) strategy (\textit{left}) and the forbidden day (FD) strategy (\textit{right}) with a zero inclination $\delta=0$. The lower panels show the exposure map with inclination angles of $-20^\circ$ (\textit{left}) and $5^\circ$ (\textit{right}) under the FI strategy.
    Some important diffuse structures in the Milky Way, including the eROSITA bubbles and Cygnus superbubbles, are marked and labeled. The position of the calibration source, Crab Pulsar, is also shown. The Crab Pulsar Nebula will not be covered by DIXE at an inclination angle of $\delta=-20^\circ$. All the maps are in ``Hammer-Aitoff" projection and Galactic coordinates.}
    \label{fig:s2_s0_inclination}
\end{figure}
DIXE’s pointing will remain fixed with respect to CSS during the survey. The DIXE pointing inclination angle, $\delta$, shown in Figure~\ref{fig:sunavd}, is defined as the angle between the pointing vector $\vec{n}_{\rm DX}$ and the CSS position vector $\vec{r}_{\rm CSS}$, and it determines the observable sky. Without loss of generality, the azimuthal angle of $\vec{n}_{\rm DX}$ in the comoving frame is set to 0. Therefore, throughout this work, we assume $\vec{n}_{\rm DX}$ lies in the $y'z'$ plane.

We consider inclination angles ranging from $5^\circ$ to $-20^\circ$, in $-5^\circ$ steps. One-year exposure maps generated using the “forbidden orbit” closing strategy (see Section~\ref{sec:sun_avoid}) are shown in Figure~\ref{fig:s2_s0_inclination}. Due to the CSS orbit's $\sim41.5^\circ$ inclination and a fixed $\delta$, a fraction of the sky will be inaccessible for all cases, shown as the ``blue holes" in Figure~\ref{fig:s2_s0_inclination}. For positive inclination angles, a significant portion of the western (particularly southwestern) eROSITA bubble region and the Vela supernova remnant becomes inaccessible. The uncovered area of the eROSITA bubbles increases as the absolute value of $\delta$ increases. Since the eROSITA bubbles are a major scientific objective for DIXE, positive inclination angles are excluded from further consideration. 

On the other hand, the Crab Nebula is a key calibration target, and inclination angles of $\delta \lesssim 20^\circ$ would exclude it from view (Figure~\ref{fig:s2_s0_inclination}). Furthermore, decreasing the inclination angle also results in more of the Cygnus superbubble region becoming inaccessible. Table~\ref{tab:skycover} summarizes the visibility of DIXE’s major science targets and the sky coverage at different inclination angles.

Taking all these factors into account, we adopt an inclination angle of $\delta = 0^\circ$, where the pointing vector aligns with the CSS position vector during the survey. The sky coverage at $\delta=0$ is $\sim 72.5\%$.  At this configuration, DIXE covers most of the eROSITA bubbles, excluding a southwestern segment ($l \sim 353^\circ$–$330^\circ$) and a northwestern region near the Galactic plane ($l \gtrsim 338^\circ$, $b \lesssim 15^\circ$), as shown in Figure~\ref{fig:s2_s0_inclination}. DIXE will observe a large portion of the Cygnus superbubble and Cygnus loop, though its eastern arc will remain unexposed. The east part of the Vela SNR and the entirety of other bubbles, such as the ERI-ORI Bubble \citep{Burros_1993_EriOriBubble, Joubaud_2019_EriOriBubble} and Monogem Ring \citep{Knies_2018_Monogem, Knies_2024_Monogem}, will also be covered. Unfortunately, both LMC and SMC fall outside the observable region.

\begin{table}
    \centering
    \caption{The visibility of the major targets and total sky covering fraction of DIXE at different inclination angles $\delta$. The asterisks indicate partial coverage of the science targets.}
    \begin{tabular}{lcccccc}
    \hline
    Inclination angle & $5$ & $0$ & $-5$ & $-10$ & $-15$ & $-20$ \\
    (deg) & & & & & & \\
    \hline
    eROSITA bubble & $\surd^*$ & $\surd^*$ & $\surd^*$ & $\surd^*$ & $\surd^*$ & $\surd^*$ \\
    Cygnus Superbubble & $\surd^*$ & $\surd^*$ & $\surd^*$ & $\surd^*$ & $\surd^*$ & $\times$ \\
    Vela SNR & $\times$ & $\surd^*$ & $\surd^*$ & $\surd^*$ & $\surd$ & $\surd$ \\
    Crab & $\surd$ & $\surd$ & $\surd$ & $\surd$ & $\surd$ & $\times$ \\
    Sky coverage (\%) & 72.2 & 72.5 & 72.2 & 71.4 & 70.0 & 68.1 \\
    \hline
    \end{tabular}
    \label{tab:skycover}
\end{table}

\subsection{Sun avoidance strategies}
\label{sec:sun_avoid}
During the survey, the Sun would periodically enter the field of view (FoV) of DIXE. Solar radiation can significantly raise the temperature of the superconducting detectors, potentially compromising their performance. Also, as a bright X-ray target, the Sun can saturate the detectors. To mitigate these risks, DIXE is designed to close a sunshade whenever the Sun enters the FoV.


We define an angular exclusion region centered on the Sun with a radius of $20^\circ$, referred to as the \textit{avoidance zone} (Figure~\ref{fig:sunavd}, \textit{right} panel). The sunshade is triggered to close when the DIXE FoV intersects this region. We define the Sun angle, $\theta_{\odot}(t)$, as the angular separation between the Sun and the center of the FoV (Figure~\ref{fig:sunavd}). This angle is a function of time due to the orbital motion of DIXE. As the Sun approaches the avoidance zone, $\theta_\odot(t)$ decreases. We define the \textit{Sun avoidance angle} as $\theta_{\rm avd} = 20^\circ + \rm FoV/2$, within which solar intrusion is considered to occur. For an FoV of $10^\circ$, the Sun avoidance angle corresponds to $\theta_{\rm avd} = 25^\circ$.

\begin{figure}
    \centering
    \includegraphics[width=\hsize,trim={1cm 0cm 1cm 0cm}]{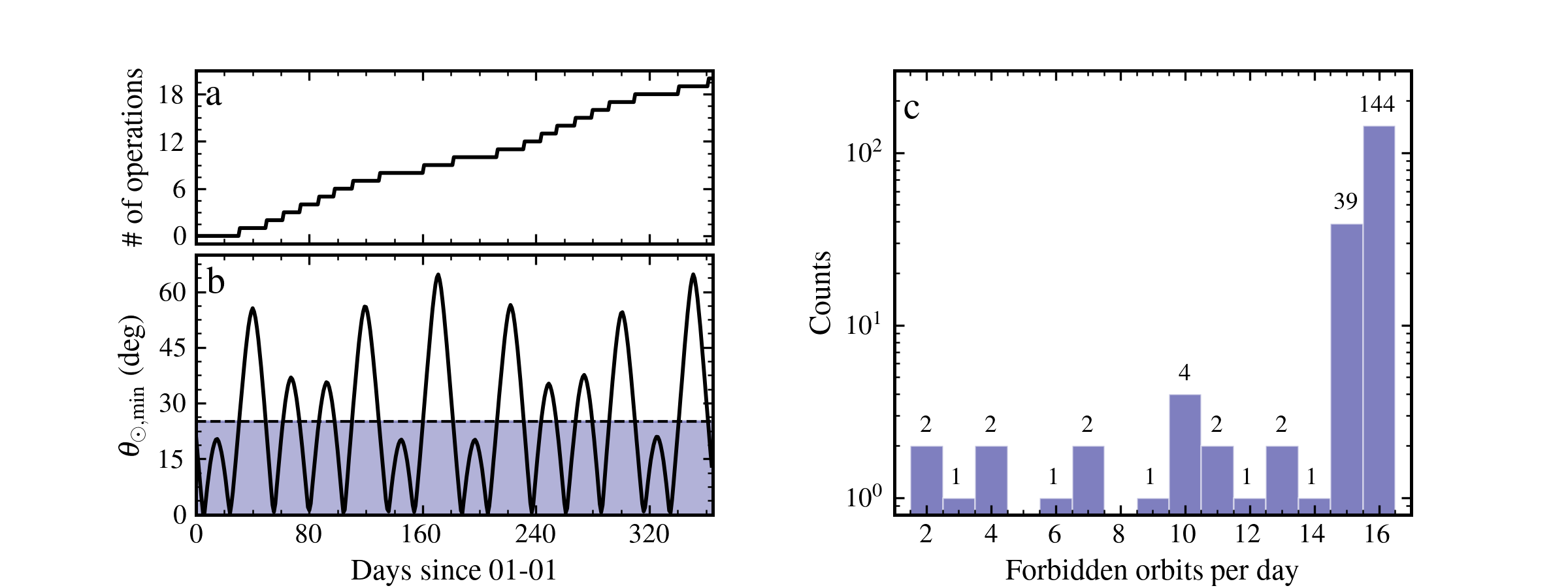}
    \caption{Panel (\textit{a}): Number of cumulative operation time since the first day of a year under the FD strategy. Panel (\textit{b}): The minimum Sun angles $\theta_{\odot,\min}$ in degrees for each day of the year. The critical angle $25^\circ$ is marked by the black dashed line. The purple-shaded region indicates days with minimum Sun angles smaller than the critical $\theta_{\rm avd}=25^\circ$. Panel (\textit{c}): Histogram of the number of forbidden orbits per day.}
    \label{fig:forbidden_days}
\end{figure}

Figure~\ref{fig:forbidden_days} shows the daily minimum value of $\theta_\odot(t)$ over a year (from 01 -- 01 to 12 -- 31). Based on the value of $\theta_\odot (t)$, we define the \textit{forbidden interval}, \textit{forbidden orbit}, and \textit{forbidden day} as follows:
\begin{itemize}
    \item \textit{forbidden interval}: the time (and also space) interval during which the FoV overlaps with the avoidance zone, i.e., when $\theta_{\odot}(t) \leq\theta_{\rm avd} =25^\circ$.
    \item \textit{forbidden orbit}: any orbit in which the Sun avoidance criterion is violated at some time, i.e., minimum $\theta_\odot (t)\le \theta_{\rm avd}=25^\circ$. The forbidden interval occurs for each forbidden orbit.
    \item \textit{forbidden day}: any day in which the Sun avoidance angle is violated at some time—i.e., the minimum daily $\theta_\odot(t)\leq\theta_{\rm avd}=25^\circ$. At least one orbit during the forbidden day will intersect with the avoidance zone.
\end{itemize}



Ideally, the sunshade should remain closed only during the forbidden interval (FI) and reopen afterward to maximize the observation duty cycle. However, this approach requires frequent mechanical operations, which could pose risks in the space environment. To mitigate this concern, we propose an alternative—the forbidden day (FD) strategy—where the sunshade remains closed throughout the entire \textit{forbidden day}. As shown in Figure~\ref{fig:forbidden_days}a, this reduces the number of sunshade operations from 5975 to just 21 over the course of a year. Figure~\ref{fig:forbidden_days}b shows that the minimum Sun angle $\theta_{\odot,\min}$ remains below the critical $25^\circ$ threshold for continuous periods, indicating that the orbit continuously intersects with the avoidance zone for several consecutive days. We further examined the number of forbidden orbits contained in a forbidden day, and the distribution is shown in Figure~\ref{fig:forbidden_days}c. In 144 out of 202 forbidden days, every orbit intersects the avoidance zone. From this perspective, the forbidden orbit-based closing strategy would provide no significant benefit over the FD strategy. 


\begin{figure}
    \centering
    \includegraphics[width=\hsize]{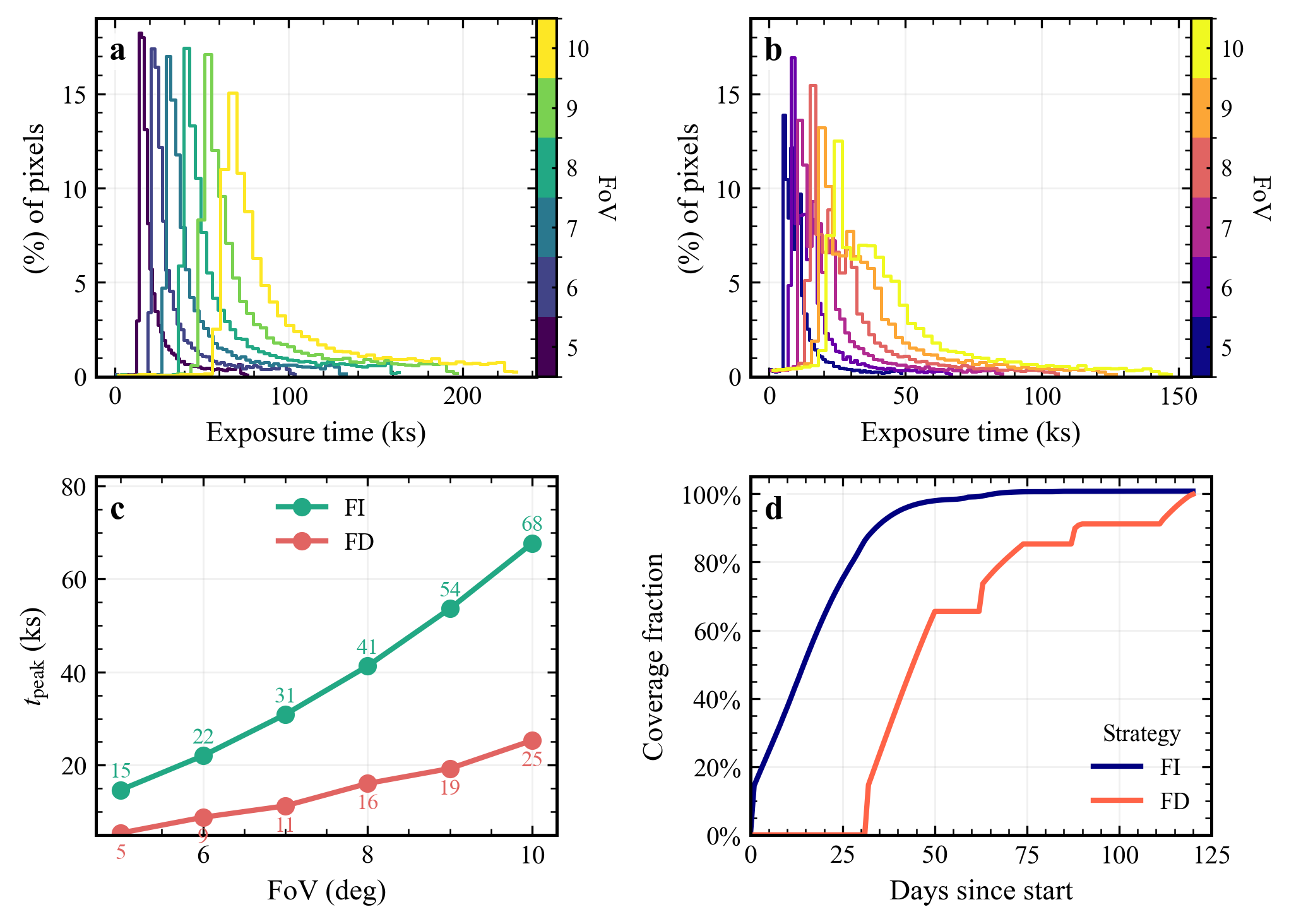}
    \caption{Comparison of the exposure-time distribution and survey efficiency for different FoV size scales (see Section~\ref{sec:fov}) under the two Sun-avoidance strategies. Panels (a) and (b) show the sky-area fraction as a function of exposure time for the forbidden-interval (FI) and forbidden-day (FD) strategies, respectively, with different FoV size scales indicated by the discrete color bars. Panel (c) shows the corresponding peak exposure time, ($t_{\rm peak}$), as a function of FoV size scale for both strategies. Panel (d) shows the evolution of the sky-coverage fraction with survey time for the FI and FD strategies.}
    \label{fig:s0_s2_expt_hist}
\end{figure}

The one-year exposure maps for the FD and FI strategies are shown in the upper panels of Figure~\ref{fig:s2_s0_inclination}, respectively. The FI strategy results in significantly higher exposure times compared to FD. Figure~\ref{fig:s0_s2_expt_hist}a -- c shows the distribution of the one-year exposure time obtained for the available sky under the FD and FI strategy. The tails at high exposure times are attributed to the orbit overlapping near the polar regions (also see Figure~\ref{fig:s2_s0_inclination}). 
We characterize the exposure map using two values: the exposure time when the histogram begins to rise, $t_{\rm rise}$, representing the minimum exposure time for most of the covered sky, with regions below this threshold contributing only marginally; And the exposure time at which the histogram peaks, $t_{\rm peak}$, representing the exposure time for the majority of the available sky. For FD and FI, the histogram peaks at $t = 26~\rm ks$ and $t = 68~\rm ks$. This indicates that for most sky regions, the FI approach provides more than twice the exposure time compared to FD. We summarize the characteristic time values in Table~\ref{tab:fov_trise_tpeak}. In conclusion, to avoid the Sun, we consider two closing strategies for DIXE's operable sunshade:
\begin{itemize}
    \item Forbidden day (FD): the sunshade is closed for an entire forbidden day. It minimizes the number of operations at the expense of exposure time.
    \item Forbidden interval (FI): the sunshade is closed only during a forbidden time interval. It maximizes observational efficiency but increases mechanical risks.
\end{itemize}

\subsection{Survey period}
\label{sect:expmap}
We compute the time required for DIXE to cover the full accessible sky. Figure~\ref{fig:s0_s2_expt_hist}d shows the sky coverage fraction as a function of time for the FD and FI strategies, assuming an observation start at the first day of the year. The required survey duration is approximately 120 and 70 days for the FD and FI strategies, respectively. Tests with different observation start dates show only minor variations in the FD survey duration, caused by seasonal changes in the Sun’s position and the resulting interruptions of the observing sequence, while the FI strategy remains nearly unchanged. The FI strategy provides both higher exposure efficiency and faster sky coverage than the FD strategy.

\subsection{Field-of-View Optimization}
\label{sec:fov}
The $10^\circ\times10^\circ$ FoV of DIXE is designed to optimize spectroscopic outcomes of diffuse Galactic sources. A larger FoV is not preferred, as for the Galactic diffuse sources, the typical temperature variation scale and the scale for emitting plasma to be physically connected are approximately $10^\circ$ \citep{Kaaret_2019_HaloSat, Qu_2024_XMM_MW}. A smaller FoV can provide finer spatial information, but at the cost of reduced total exposure time and thus poorer photon statistics.
To quantify the trade-off, we computed one-year exposure maps for FoVs of $5^\circ\times5^\circ$, $8^\circ\times8^\circ$, $9^\circ\times9^\circ$, and $10^\circ\times10^\circ$, each evaluated under the two Sun-avoidance strategies (FD and FI) described in Section~\ref{sec:sun_avoid}.

The characteristic exposure times are summarized in Table~\ref{tab:fov_trise_tpeak} and Figure~\ref{fig:s0_s2_expt_hist}c. Overall, increasing the FoV shifts the exposure distribution toward higher ($t_{\rm peak}$), as a larger overlapping area between adjacent scans results in more accumulated exposure. The main benefit of a larger FoV is the improved spectral quality, especially for low-surface-brightness structures
As an example, consider a CGM component along a line of sight with a temperature of $\sim 0.22~\rm keV$ and an intrinsic surface brightness of $S \sim 0.4 \times 10^{-12}\rm ~erg~s^{-1}~deg^{-2}$ in the $0.5–2~\rm keV$ band \citep{Henley_2013_XMM_MWH}, corresponding to a photon surface brightness of $S_{\rm photon} \sim 4.6 \times 10^{-4}\rm ~photons~s^{-1}~cm^{-2}~deg^{-2}$ \citep{Henley_2013_XMM_MWH}. Under these conditions, a $10^\circ\times10^\circ$ FoV enables DIXE to collect approximately 3000 photons (about 12 photons per energy bin for a $6\,\rm eV$ bin width) over a one-year survey.
Therefore, a $10^\circ\times10^\circ$ FoV is adopted as the baseline design for DIXE.

\begin{table}
  \centering
  \caption{The rising exposure time $t_{\rm rise}$ and peak exposure time $t_{\rm peak}$ for different FoVs under the two Sun-avoidance
  strategies.}
  \label{tab:fov_trise_tpeak}
  \begin{tabular}{lccccccc}
  \hline
  \hline
  \multirow{2}{*}{Strategy} & {Exposure Time} & \multicolumn{6}{c}{FoV (deg$\times$deg)} \\
  \cline{3-8}
   & (ks)  & $5\times5$ & $6\times6$ & $7\times7$ & $8\times8$ & $9\times9$ & $10\times10$ \\
  \hline
  \multirow{2}{*}{FD} & $t_{\rm rise}$ & 4  & 8  & 12 & 12 & 14 & 17 \\
   & $t_{\rm peak}$ & 5  & 9  & 12 & 16 & 20 & 26 \\
  \multirow{2}{*}{FI} & $t_{\rm rise}$ & 13 & 20 & 28 & 38 & 50 & 58 \\
   & $t_{\rm peak}$ & 16 & 24 & 34 & 42 & 54 & 68 \\
  \hline
  \end{tabular}
  \end{table}



\section{Effective Area}

\begin{figure}
    \centering
    \includegraphics[width=0.6\linewidth]{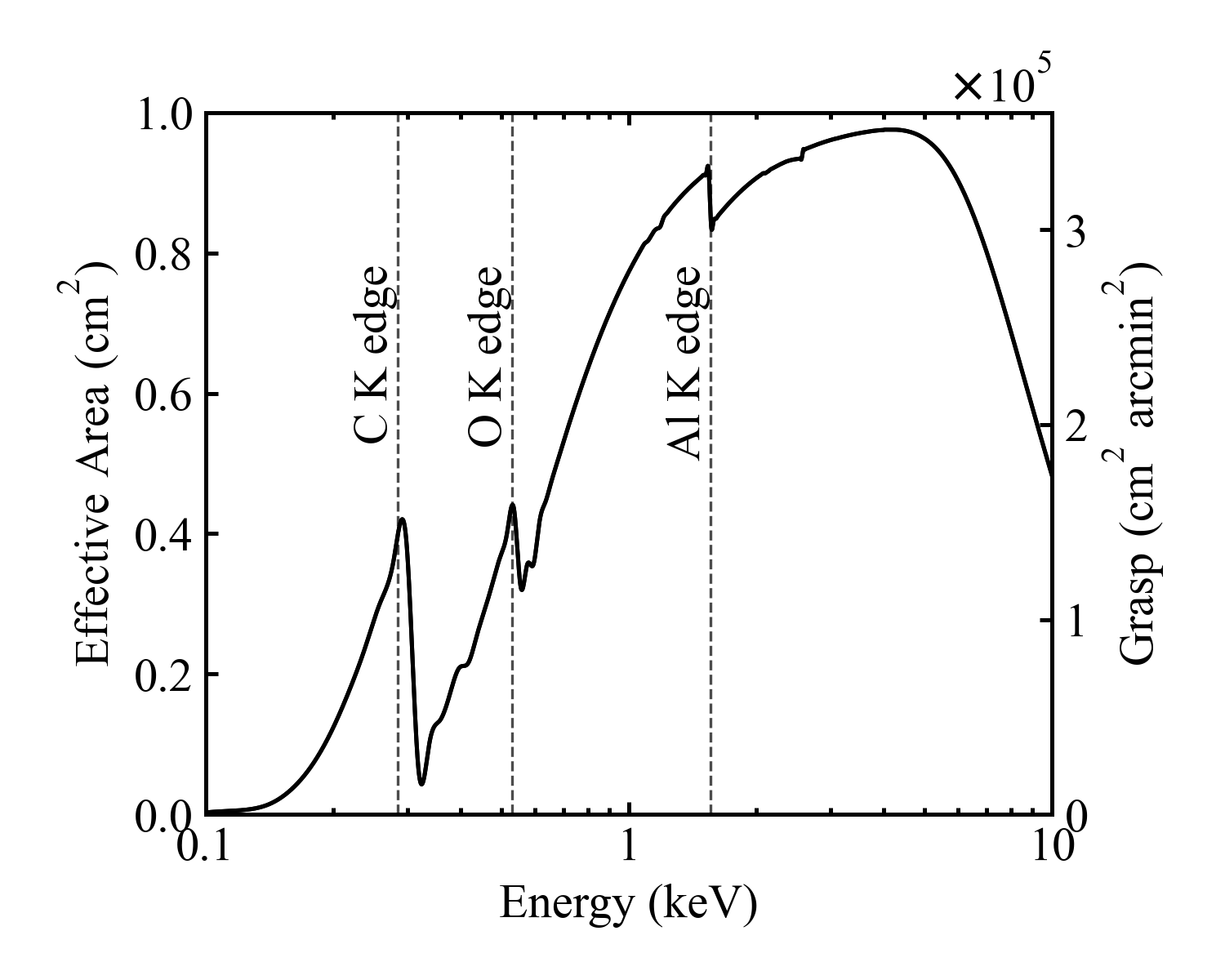}
    \caption{Effective area ($A_{\rm eff}$, left axis) and grasp (${\rm FoV}\times A_{\rm eff}$ right axis) of DIXE as a function of energy. The effective area accounts for the $1~\mathrm{cm^2}$ detector geometry, filter transmission (6 nm \ce{Al2O3}/14 Al nm/100 nm polyimide stack), and quantum efficiency (0.1 $\mathrm{\mu m}$ Au/5 $\mathrm{\mu m}$ Bi TES absorber with 99\% filling factor). The grasp represents the product of the effective area and the field of view. Energy ranges from 0.1 to 10 keV with logarithmic scaling. Prominent absorption edges contributed by filter materials are marked and labeled.}
    \label{fig:effarea_and_grasp}
\end{figure}

For a collimated telescope, the effective area is determined by the detector geometric area $A_{\rm chip}$, the filter transmission, and the detector quantum efficiency (QE). The DIXE detector is designed with an active area of $1~\mathrm{cm^2}$. Optical-blocking filters are employed to block UV/optical/IR photons, with the preliminary design consisting of aluminum-coated polyimides: 20 nm aluminum and 100 nm polyimide substrate \citep{2024JLTP..tmp..106L}, taking the 30\% of aluminum oxidation into consideration. The QE is primarily set by the absorber thickness of the TES microcalorimeter, which comprises 0.1 $\rm \mu m$ of gold (Au) and 5 $\rm \mu m$ of bismuth (Bi). A pixel filling factor of 99\% is assumed to account for the gaps between adjacent pixels. Incorporating all these factors, the resulting effective area curve is shown in Figure~\ref{fig:effarea_and_grasp} (left axis), with the corresponding grasp—defined as the product of FoV and effective area—shown on the right axis.

\section{Demodulation of collimator image}
\label{sec:collimator_response}
Although the spatial resolution is restricted by the FoV, with the tracks of DIXE's position and collimator response, an algorithmic method can be implemented to improve the imaging quality. We will introduce DIXE collimator response (hereafter CR) and the collimator response track (CRT) in this section. Based on the CRTs, we introduce the method to constrain the position or surface brightness distributions for point-like and diffuse targets.

\subsection{Collimator response}
\begin{figure}
    \centering
    \includegraphics[width=0.9\hsize]{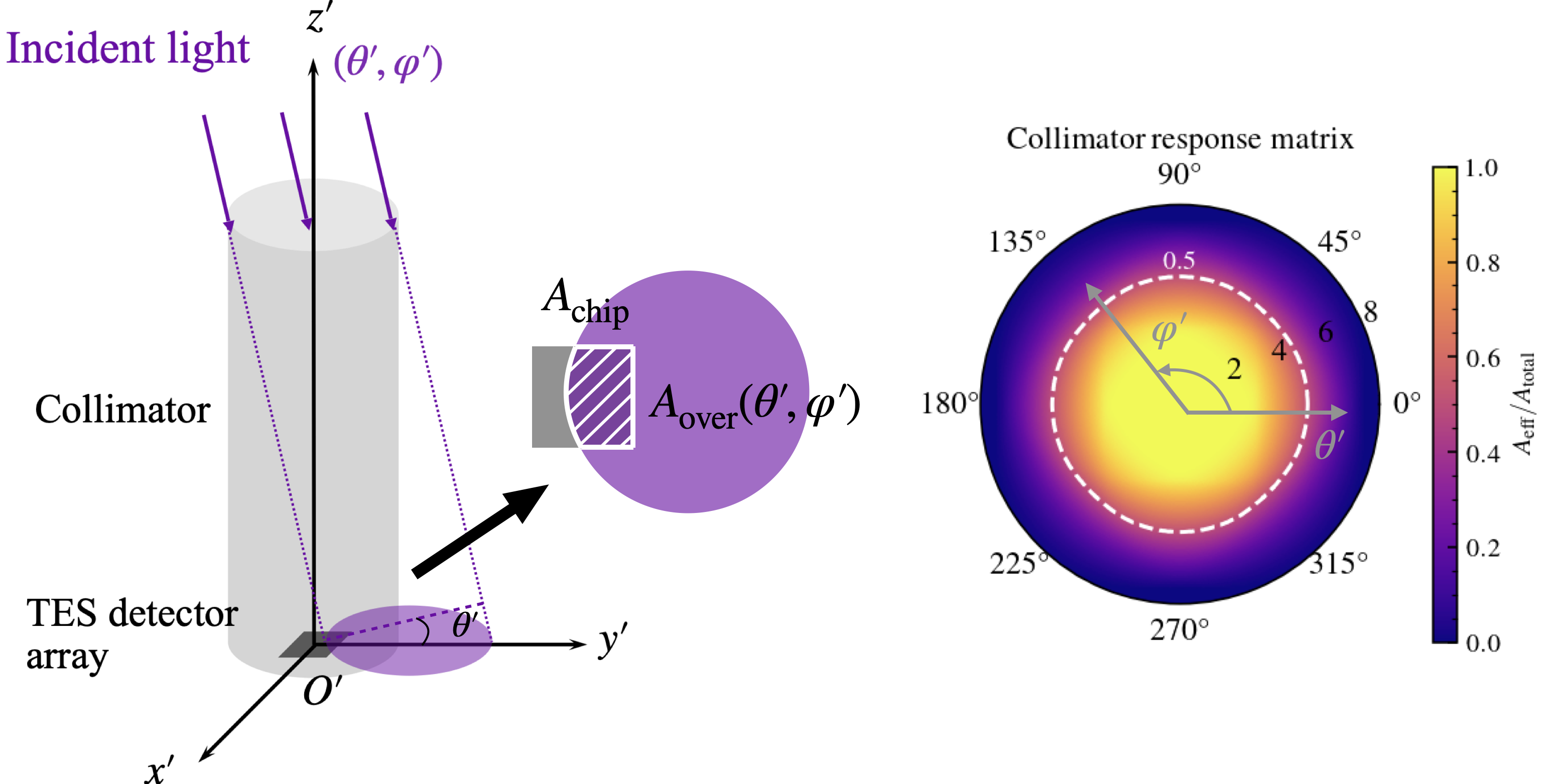}
    \caption{(\textit{Left}) Schematic illustration of the DIXE collimator geometry. Incident photons enter the cylindrical collimator with detector-frame zenith and azimuth angles $(\theta',\varphi')$, where the collimator axis is aligned with the detector normal.
    (\textit{Middle}) Projection onto the detector plane. The square represents the detector chip, i.e., the TES array footprint, and the purple circle represents the projected collimator opening. Their overlapping area, $A_{\rm over}(\theta',\varphi')$, determines the effective collecting area for a photon incident from $(\theta',\varphi')$ and forms the basis of Equation~(1). For on-axis incidence the projected opening is centered on the chip, while for off-axis incidence it is shifted on the detector plane.
    (\textit{Right}) Polar map of the normalized collimator response, color-coded by amplitude. The polar radius and angle correspond to the detector-frame zenith angle $\theta'$ and azimuthal angle $\varphi'$, respectively. The dashed white curve marks the response level of 0.5, which is used to define the half-power diameter of the field of view. The mild azimuthal asymmetry arises from the square detector geometry.}

    \label{fig:cr_phase_space}
\end{figure}

The collimator response (CR) is defined as
\begin{equation}
\mathrm{CR}(\theta',\varphi') = \frac{A_{\rm over,\perp}(\theta',\varphi')}{A_{\rm chip}}=\frac{A_{\rm over}(\theta',\varphi')\cos\theta'}{A_{\rm chip}},
\end{equation}
where $A_{\rm over,\perp}$ is the component of the overlap area perpendicular to the incident light direction, and $A_{\rm chip}$ is the total detector area (Figure~\ref{fig:cr_phase_space}, \textit{left} panel). The angles $\theta'$ and $\varphi'$ denote the incident zenith and azimuthal angles in the detector frame (indicated by primes throughout this paper).
The detector coordinate system $(x',y',z')$ is defined with the origin ($O'$) at the center of the detector chip. The $x'O'y'$ plane coincides with the detector plane, the $z'$ axis is aligned with the nominal pointing direction (i.e., the DIXE positional vector at zero inclination), and the $x'$ and $y'$ axes are aligned with the orbital tangential and normal directions, respectively. The detector boundaries are assumed to be parallel to the $x'$ and $y'$ axes.

The right panel of Figure~\ref{fig:cr_phase_space} shows the CR varied with photon incident angle $\theta'$ and $\varphi'$ in a polar projection. The polar radius and polar angle indicate the values of $\theta'$ and $\varphi'$. A slight asymmetry with respect to $\varphi'$ is observed due to the square geometry of the detector. The CR value declines with increasing $\theta'$. The FoV, defined as the full-width half maximum (FWHM) of the CR profile, is $\sim 10^\circ$.

\subsection{collimator response track and demodulation method based on the Monte-Carlo Markov Chain}
\subsubsection{Point-like sources}
\begin{figure}
    \centering
    \includegraphics[width=\hsize,trim={2cm 0cm 2cm 0cm}]{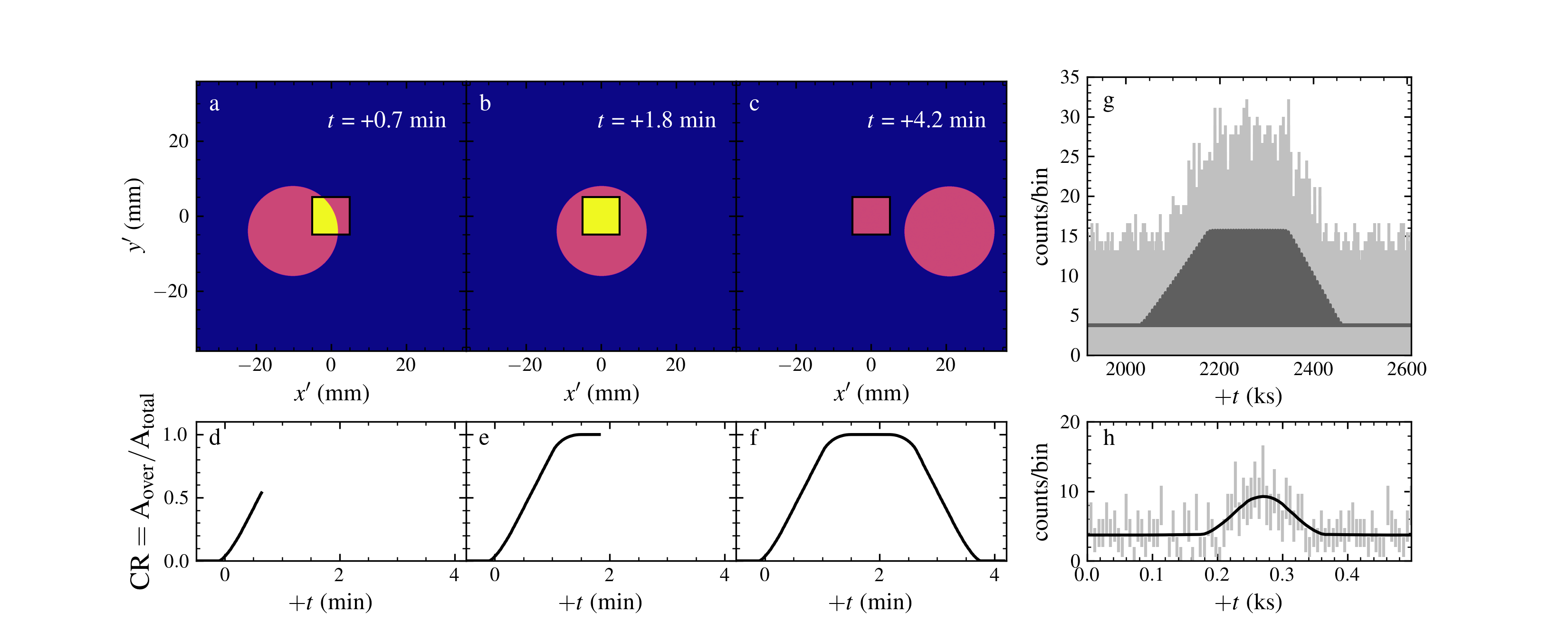}
    \caption{Panel \textit{a -- c}: the projected circular entrance of the collimator in the detector ($x'O'y'$) plane at different time slots since the beginning of the simulation. Panel \textit{d -- f}: the collimator response varied with time since the beginning of the simulation. Panel \textit{g} shows the mocked counts per bin for an ideal point-like source at $(\rm ra,dec)=(150,25)$ for 8 days. The source flux are assumed to be $2~\rm photons~s^{-1}~cm^{-2}$ and the background count rate is set to $0.59~\rm counts~s^{-1}$. The mocked data with errors is shown in silver, and the model is shown in dark gray. Panel \textit{h} shows the zoomed-in view of panel \textit{g} during a single scan, where the data is shown by silver dashes and the model is shown by solid black curves.}
    \label{fig:cr_tracks}
\end{figure}
During the sky survey, the DIXE pointing continuously evolves, causing a source at fixed celestial coordinates to repeatedly enter and exit the field of view. For a point source located at $({\rm ra}, {\rm dec})$, the photon incident angles $(\theta',\varphi')$ in the detector frame vary with time as DIXE scans the sky. The collimator response can therefore be written as
\begin{equation}
{\rm CR}({\rm ra}, {\rm dec}, t) = \frac{A_{\rm over,\perp}\left[\theta'({\rm ra}, {\rm dec}, t), \varphi'({\rm ra}, {\rm dec}, t)\right]}{A_{\rm chip}},
\label{eq:crtrack}
\end{equation}
which explicitly depends on time through the incident direction. We refer to this time-dependent collimator response as the \emph{collimator response track} (CRT).

DIXE orbits around the Earth about 16 times per day. During a single orbit, the projected circular entrance of the collimator (see Figure~\ref{fig:cr_phase_space}) in the detector plane moves in the $x'$-axis direction. This process is illustrated in 
Figure~\ref{fig:cr_tracks}, panels \textit{a -- c}, and the corresponding CRTs are shown in panels \textit{d -- f}. The CR value rises as the source enters DIXE FoV (panel \textit{a}) and reaches the peak when the incident light illuminates the whole detector chip (panel \textit{b}) and declines as the source exits the FoV (panel \textit{c})

As shown by Eq.~\ref{eq:crtrack}, the CRT is essentially a time-dependent correction factor of the effective area at a given exposure time, and is determined by the position of the source. 
Thus, considering an ideal point source, within a given time bin $\Delta t$, the count rate observed by DIXE is:
\begin{equation}
    {\rm counts/bin} =\Delta t\times\left[\tilde f_{\rm src} \times \mathrm{CR}(t; {\rm ra}, {\rm dec})\times A_{\rm chip} + \tilde F_{\rm bkg} \right]
    \label{eq:cts_per_bin}
\end{equation}
Where $\tilde f_{\rm src}$ is the observed source flux in units of $\rm photons~s^{-1}~cm^{-2}$, representing the flux received by the detector after attenuation by the filters. $\tilde F_{\rm bkg}$ is the observed background count rate in units of $\rm counts~s^{-1}$, including contributions from both the cosmic X-ray background (CXB) and the particle (non-X-ray) background \citep[NXB;][]{Tian_2026}. Two background levels are considered and summarized in Table~\ref {tab:bkg}, with detailed calculations provided in Section~\ref{sec:bkg}. Figure~\ref{fig:cr_tracks}, panel \textit{g}, shows the simulated count rate profile for a point source located at $({\rm ra},{\rm dec})=(150^\circ, 25^\circ)$, assuming $\tilde f_{\rm src} = 2~\rm photons~s^{-1}~cm^{-2}$, $\tilde F_{\rm bkg} = 0.59~\rm counts~s^{-1}$, and a time bin size of $\Delta t = 6~\rm s$. The model count rate, computed using Eq.~\ref{eq:cts_per_bin}, is plotted as the dark gray curve. The seemingly “filled” pattern in the profile is caused by the pulse-like shape of the signal when the DIXE FoV repeatedly sweeps over the source. Panel \textit{h} presents a zoomed-in view of a single orbit segment from panel \textit{g}, illustrating the structure more clearly.

Eq.~\ref{eq:crtrack} indicates that the source, background flux, and the source coordinates determine the normalization and shape of the count rate profile. By fitting the count rate profile with free parameters $({\rm ra},{\rm dec},~\tilde f_{\rm src},~\tilde F_{\rm bkg})$, we can put a coordinate constraint on the point source. We use the method based on the Monte-Carlo Markov Chain (MCMC; \citealt{2020sdmm.book.....I}) to fit the count rate profile. The model used in MCMC, $M(\vec{\vartheta}|t)$ is shown by Eq.~\ref{eq:cts_per_bin}, 
where the parameters $\vec\vartheta=({\rm ra,dec},\tilde f_{\rm src},\tilde F_{\rm bkg})$, or
\begin{equation}
   M(\vec{\vartheta}|t) =M({\rm ra,dec},\tilde f_{\rm src},\tilde F_{\rm bkg}|t)= \Delta t \times\left[\mathrm{CR}(\mathrm{ra},\mathrm{dec}|t)\times \tilde f_{\rm src}\times A_{\rm chip} + \tilde F_{\rm bkg}\right]  
\end{equation}
The observed count rate follows a Poisson distribution, and the likelihood for all measurements is given by
\begin{equation}
    L_{p} = \sum_i\left(y_i\ln\lambda_i-\lambda_i -\ln[\Gamma(y_i+1)]\right) 
\end{equation}
where $y_i$ and $\lambda_i$ are the observed and expected counts for each bin. $\lambda_i$ is calculated by $M(\vec\vartheta,t_i)$. $\Gamma(x)$ is the gamma function, which is essentially factorial for integers. 

\begin{table}
    \centering
    \caption{Background level $\tilde F_{\rm bkg}$ including CXB and NXB received by DIXE in $\ctrate$. The $N_{\rm H}$ column listed the assumed neutral hydrogen column density. The NXB level is estimated from Hitomi measurements.}
    \begin{tabular}{cccc}
    \hline
    $N_{\rm H}$ & $\tilde F_{\rm NXB}$ & $\tilde F_{\rm CXB}$ & $\tilde F_{\rm bkg,~2-8~\rm keV}$ \\
    $(\rm cm^{-2})$ &  $(\rm counts~\rm s^{-1})$ & $(\rm counts~\rm s^{-1})$ & $(\rm counts~\rm s^{-1})$ \\
    \hline
    $3\times10^{22}$ & 0.38 & 0.15 & 0.53 \\
    $3\times10^{20}$ & 0.38 & 0.21 & 0.59 \\
    \hline 
    \end{tabular}
    \label{tab:bkg}
\end{table}

\subsubsection{Extended sources}
\begin{figure}[h]
 \begin{picture}(0,0)
 \put(0,145){\makebox(0,0)[l]{{\color{black}\textbf{a}}}}
 \put(260,145){\makebox(0,0)[l]{{\color{black}\textbf{b}}}}
 \end{picture}
  \begin{minipage}[t]{0.58\hsize}
  \centering
   \includegraphics[width=\hsize,trim={1.5cm 0cm 0cm 0cm}]{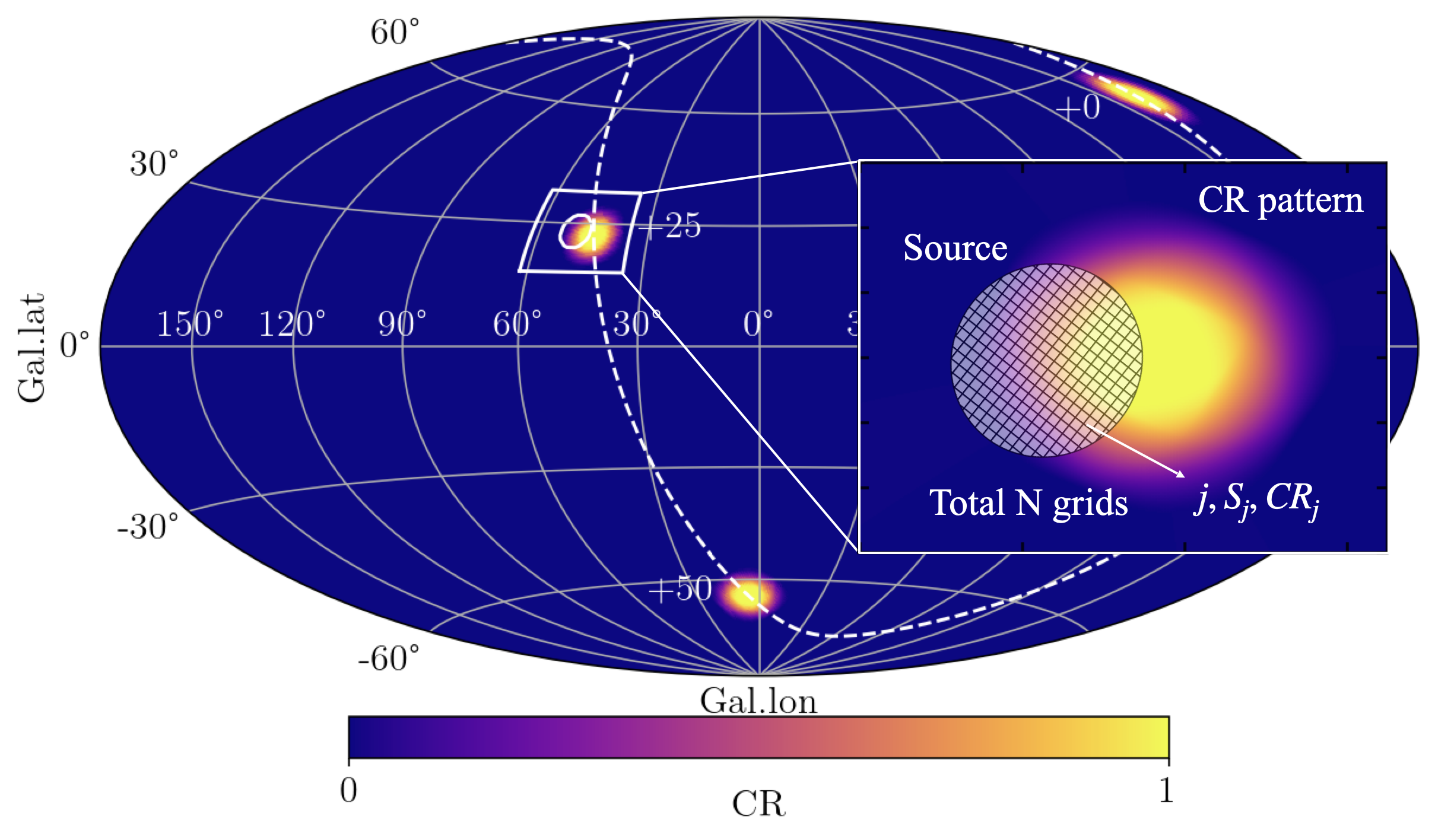}
  \end{minipage}
  \begin{minipage}[t]{0.36\hsize}
  \centering
   \includegraphics[width=\hsize]{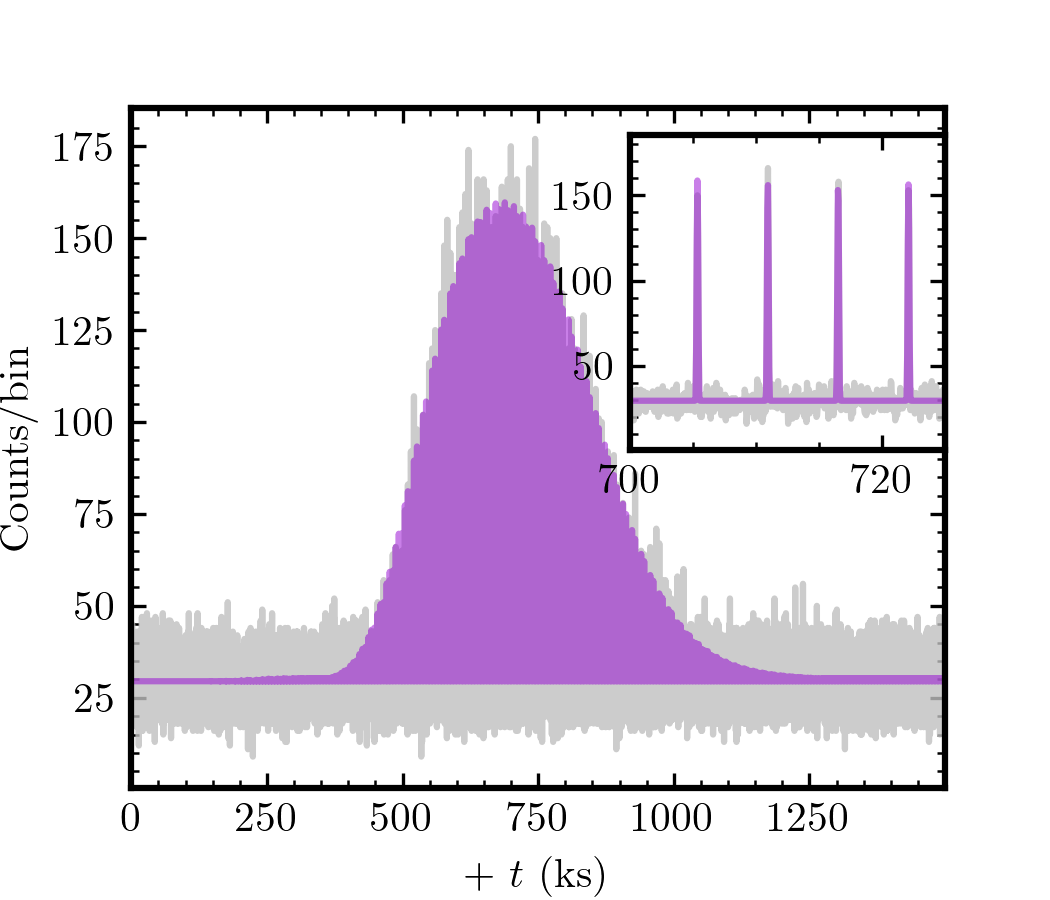}
  \end{minipage}
    \caption{(\textit{a}) Collimator response (CR) projected in Galactic coordinates at three times during one orbit: start (+0), +25 min, and +50 min. Colors denote CR values, and the dashed white line indicates the scanning trajectory. The white circle marks an extended source partially covered by the DIXE FoV. The inset shows a zoomed-in source region divided into $N$ elements with surface brightness $S_j$ and corresponding CR$_j$, used to compute the total count rate via Eq.~\eqref{eq:riemann_sum} (\textit{b}) Simulated count-rate profile for a Gaussian surface-brightness distribution with FWHM $=5^\circ$ and peak $\bar{S}_0=0.1$~photons~s$^{-1}$~cm$^{-2}$~deg$^{-2}$, including a background of $\bar{F}_{\rm bkg}=0.59$~counts~s$^{-1}$ ($2-8$~keV), integrated over $\sim$17~days. The dark violet curve shows the model prediction, and the silver curves show mock data with uncertainties. The inset highlights four individual scans.
    }

    \label{fig:psfsky_gauss_countrate}
\end{figure}
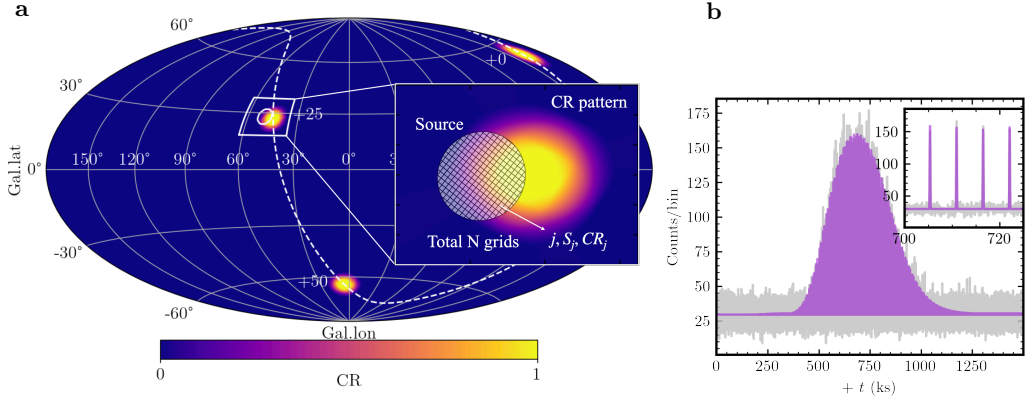

For extended sources, it is more convenient to work in the sky coordinate frame. At a given time $t$, the DIXE position and pointing direction are known, allowing each sky position in the celestial frame $\Sigma$ to be mapped to the detector frame $\Sigma'$, characterized by the incident angles $(\theta',\varphi')$ (Figure~\ref{fig:cr_phase_space}, left panel). The collimator response $\mathrm{CR}(\theta',\varphi')$ (Figure~\ref{fig:cr_phase_space}, right panel) can therefore be expressed in the sky frame as $\mathrm{CR}(\mathrm{ra},\mathrm{dec};t)$. The top panel of Figure~\ref{fig:psfsky_gauss_countrate}a illustrates the sky-projected CR pattern at three representative times. For an extended source with specific surface brightness distribution $S_E(\mathrm{ra},\mathrm{dec};E)$, the total number of detected counts $C$ accumulated over an exposure $\Delta t$ is
\begin{equation}
    C=\Delta t\,A_{\rm chip}\int \mathcal{T}(E)\,dE \int S_E(\mathrm{ra},\mathrm{dec};E)\,\mathrm{CR}(\mathrm{ra},\mathrm{dec};t)\,d\Omega,
    \label{eq:diffuse_integal}
\end{equation}
where $\mathcal{T}(E)$ is the combined dimensionless throughput factor including transmission, filling factor, and quantum efficiency ($A_{\rm eff}=\mathcal T(E)\times A_{\rm chip}$, Figure~\ref{fig:effarea_and_grasp}). $d\Omega$ is the differential solid angle. Let
\begin{equation}
    \tilde S(\mathrm{ra},\mathrm{dec})=\int S_E(\mathrm{ra},\mathrm{dec};E)\times \mathcal{T}(E)\,dE,
\end{equation}
then Equation~\ref{eq:diffuse_integal} can be rewritten as
\begin{equation}
    C=A_{\rm chip}\Delta t\int \tilde S(\mathrm{ra},\mathrm{dec}) \times \mathrm{CR}(\mathrm{ra},\mathrm{dec};t)\,d\Omega.
    \label{eq:int}
\end{equation}
As shown by the insets of Figure~\ref{fig:psfsky_gauss_countrate}a, by dividing the source region into $N$ elements, with each grid point $j$ having a surface brightness $\tilde S_j$, solid angle $\omega_j$, and corresponding ${\rm CR}_j$, Equation~\ref{eq:int} can be approximated by the Riemann summation
\begin{equation}
    C\approx A_{\rm chip}\Delta t\sum_{j=1}^N{\rm CR}_j\,\tilde S_j\,\omega_j.
    \label{eq:riemann_sum}
\end{equation}
Compared with Eq~\ref{eq:cts_per_bin}, the treatment of extended sources is formally similar to the point-source case. The input model in MCMC can be written as
\begin{equation}
    M(\vec\vartheta|t),\quad \vec\vartheta=(\vec\vartheta',\tilde F_{\rm bkg}).
\end{equation}
As an example, Figure~\ref{fig:psfsky_gauss_countrate}b shows the mocked 17-day count-rate profile of a Gaussian-like extended source. The surface brightness distribution is
\begin{equation}
    \tilde S(\vec\vartheta)=S_0\frac1{\sqrt{2\pi}\sigma}\,\exp\left(-\frac{|\vec r -\vec r_0|^2}{2\sigma^2}\right),
\end{equation}
where $|\vec r -\vec r_0|$ is the angular separation with respect to the central coordinate $\vec r_0=(\mathrm{ra_0},\mathrm{dec_0})$, and $\vec\vartheta=(S_0,\sigma,\mathrm{ra_0},\mathrm{dec_0})$. We assume $\tilde S_0=0.1~\rm \phrate~cm^{-2}~deg^{-2}$ in the $2-8~\rm keV$ band and $\mathrm{FWHM}=2\sqrt{2\ln2}\sigma=5^\circ$. The background level in the $2-8~\rm keV$ band is assumed to be $\tilde F_{\rm bkg}=0.59~\rm \ctrate$. It is noticed that the shape of the count-rate profile differs from that of a point-like source, as shown in Figure~\ref{fig:cr_tracks}. This difference arises because the DIXE FoV may not fully encompass the source at any given time due to its spatial extent. Nevertheless, MCMC fitting can also be applied to constrain the underlying surface brightness distribution.

\section{Scientific capability}
\label{sec:science_capability}
We adopt the demodulation method described in Section~\ref{sec:collimator_response} and apply it to targets representing different science cases. In this section, we present the results and assess the imaging capability of DIXE. The capability of locating unanticipated point-like X-ray transients, considered bonus science for DIXE as a survey mission, is demonstrated in Section~\ref{imaging:ps}. In Section~\ref{imaging:diffuse}, we explore the reconstruction of diffuse Galactic structures with enhanced angular resolution. 
\subsection{Point source localization}
\label{imaging:ps}
\begin{figure}
    \centering
    \includegraphics[width=\hsize]{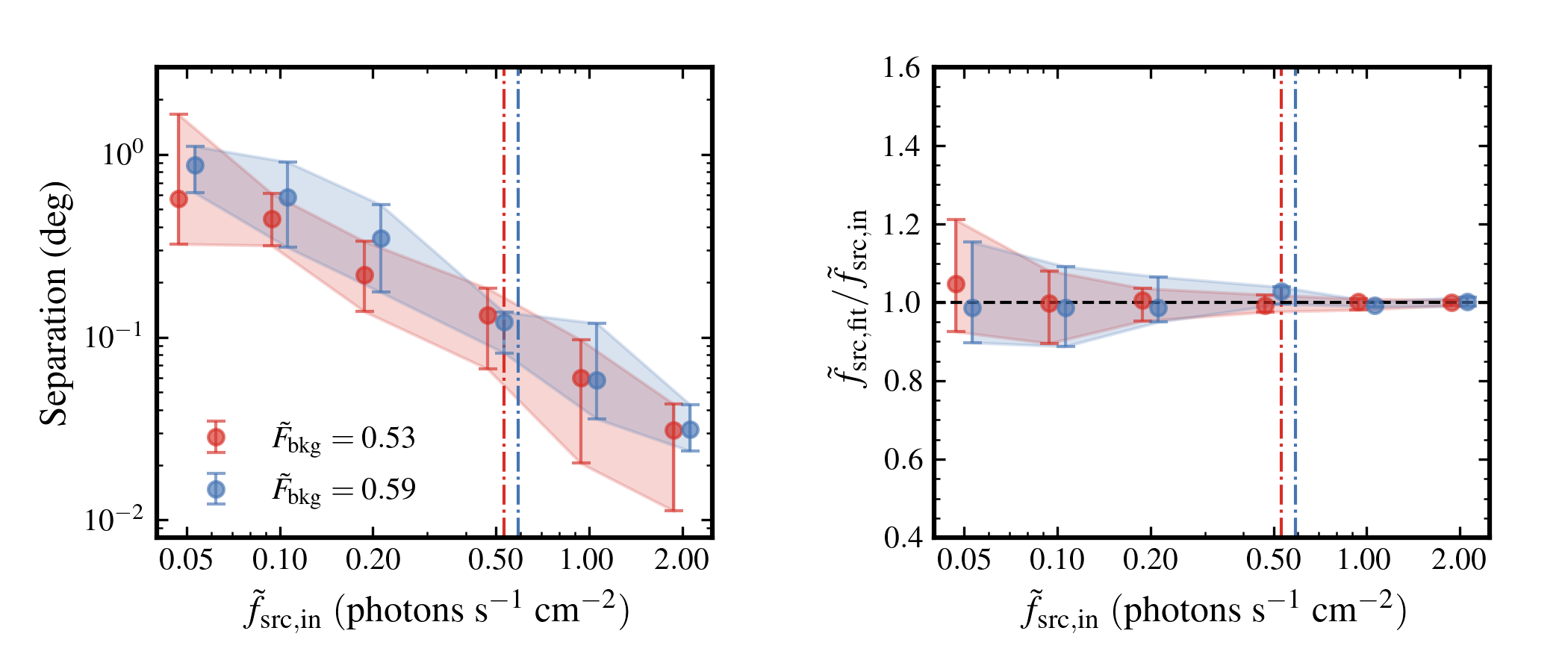}
    \caption{(\textit{Left}) Angular separation between the best-fit and input source positions as a function of the input source flux $\bar{f}_{\rm src,in}$. Results for different background assumptions (Table~\ref{tab:bkg}) are shown in different colors. 
    (\textit{Right}) Ratio of the best-fit to input source flux, $\bar{f}_{\rm src,fit}/\bar{f}_{\rm src,in}$, versus $\bar{f}_{\rm src,in}$. The horizontal dashed line marks unity. Vertical lines indicate the corresponding background flux levels ($\bar{F}_{\rm bkg}/A_{\rm chip}$ with $A_{\rm chip}=1$~cm$^{2}$). Error bars and shaded bands denote $1\sigma$ uncertainties in both panels.
    }

    \label{fig:ps_loc_mcmc}
\end{figure}

For different background levels listed in Table~\ref{tab:bkg}, we assume point sources located at $({\rm ra}, {\rm dec}) = (150^\circ, 30^\circ)$ with fluxes $f_{\rm src}$ ranging from $0.05$ to $2~\rm photons~s^{-1}$. We fit the count rate profiles over 17 days. Although uncertainties depend on the exposure time and transient fluxes may vary on multi-day timescales, here we focus on a simplified case with a full 17 days of data to demonstrate the method’s feasibility. 
Figure~\ref{fig:ps_mcmc_corner} shows an example corner plot in the parameter space (ra, dec, $\tilde f_{\rm src}$, $\tilde F_{\rm bkg}$) for source and background count rate of $0.1~\phrate~cm^{-2}$ and $0.59~\ctrate$ in $2-8~\rm keV$.
The \textit{left} panel of Figure \ref{fig:ps_loc_mcmc} shows the resulting positional accuracy for a source at $({\rm ra}, {\rm dec}) = (150^\circ, 30^\circ)$, with the observed count rate $\tilde f_{\rm src}$ ranging from $0.05$ to $2.0~\phrate~\rm cm^{-2}$. Results for different assumed background levels (from Table~\ref{tab:bkg}) are also presented.

A clear degradation in positional accuracy is observed as the source flux decreases, which is expected since a higher flux leads to a better signal-to-noise ratio and improves the fitting performance.
Across all tested conditions, our method constrains the source position to within $\sim1^\circ$ for sources with observed flux $\tilde f_{\rm src} > 0.1~\phrate~\rm cm^{-2}$ in $2-8~\rm keV$—a substantial improvement over the $10^\circ$ collimator FWHM.

To estimate the intrinsic source flux without convolving DIXE’s effective area, we assume hard spectra for the transients, where most of the flux lies in the $2-8~\rm keV$ band. Since the effective area remains close to unity in the $2-8~\rm keV$ range, the intrinsic flux in $2-8~\rm keV$ is approximately equal to the observed value, i.e., $f_{\rm src} \sim 0.1~\phrate~\rm cm^{-2}$, corresponding to roughly 30 mCrab. The method fails to converge when the source flux $f_{\rm src}\lesssim 0.2~\rm \ctrate$, and is therefore only applicable to relatively bright sources.

The \textit{right} panel of Figure~\ref{fig:ps_loc_mcmc} shows the constraints on the source flux. For sources with $\tilde f_{\rm src} > 0.1~\rm \phrate~cm^{-2}$ in $2-8~\rm keV$, the ratio between the best-fit and input flux indicates a deviation of $\lesssim 20\%$. The fitting accuracy improves progressively with increasing source flux.

\begin{figure}
    \centering
    \includegraphics[width=0.9\hsize]{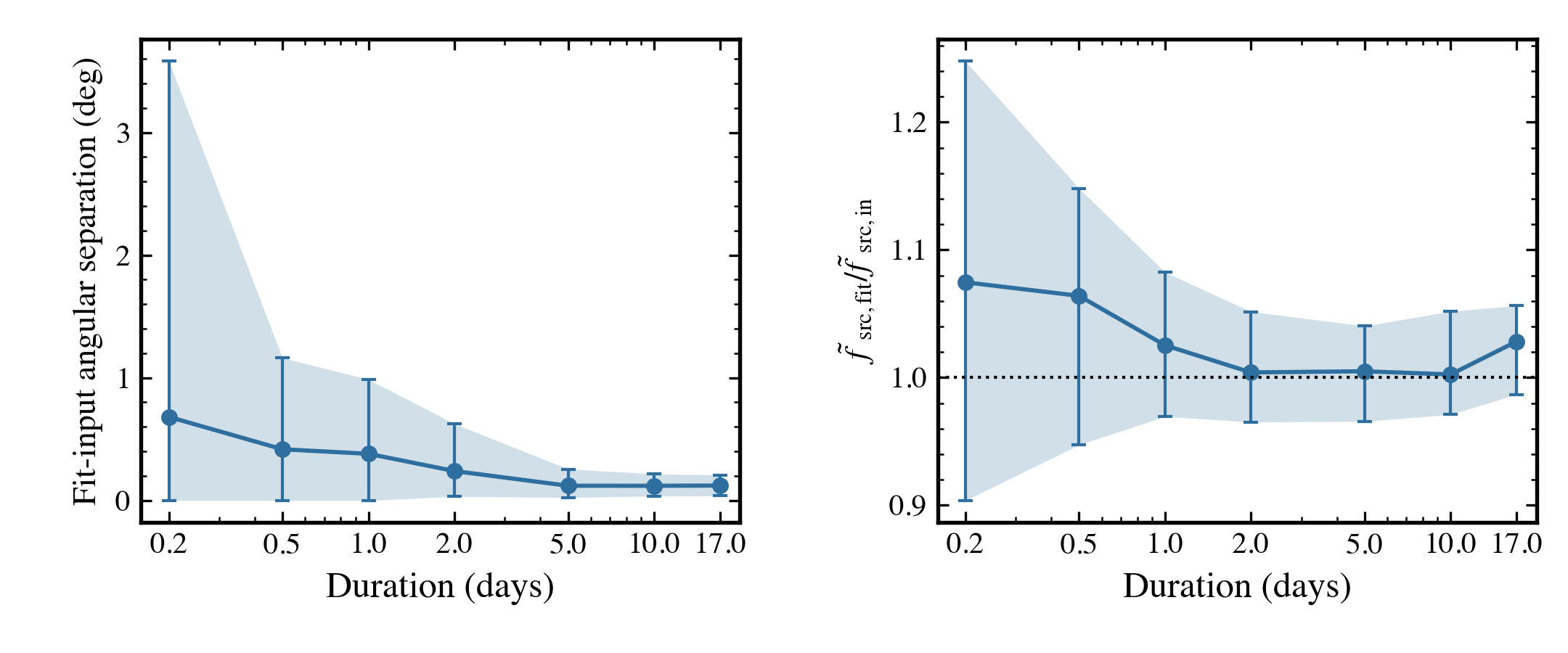}
    \caption{Point-source fitting performance as a function of fitted duration for an ideal point source with $\tilde f_{\rm src}=0.5~\phrate$ and $\tilde F_{\rm bkg}=0.59~\ctrate$. The left panel shows the angular separation between the fitted and input source positions, and the right panel shows the recovered source flux normalized by the input value. Points and shaded regions show the median and central 68\% scatter over 10 Poisson realizations.}
    \label{fig:ps_duration}
\end{figure}

The variability of the source flux is not considered yet. For sufficiently high signal-to-noise ratios and variability timescales longer than about 0.5 day, the source coordinates are constrained mainly by the timing of the rise and decline of the CRT, whereas the source flux mainly affects the modulation amplitude. Therefore, variability is expected to broaden the flux uncertainty more strongly than the positional uncertainty, provided that the selected interval contains an effective scan across the source. A duration test for a representative bright source confirms that the median localization remains within $\sim1^\circ$ once the fitted interval is $\gtrsim0.5$ day (Figure~\ref{fig:ps_duration}).

 \begin{figure}
      \centering
      \includegraphics[width=\hsize,trim={0.2cm 0cm 1cm 0cm}]{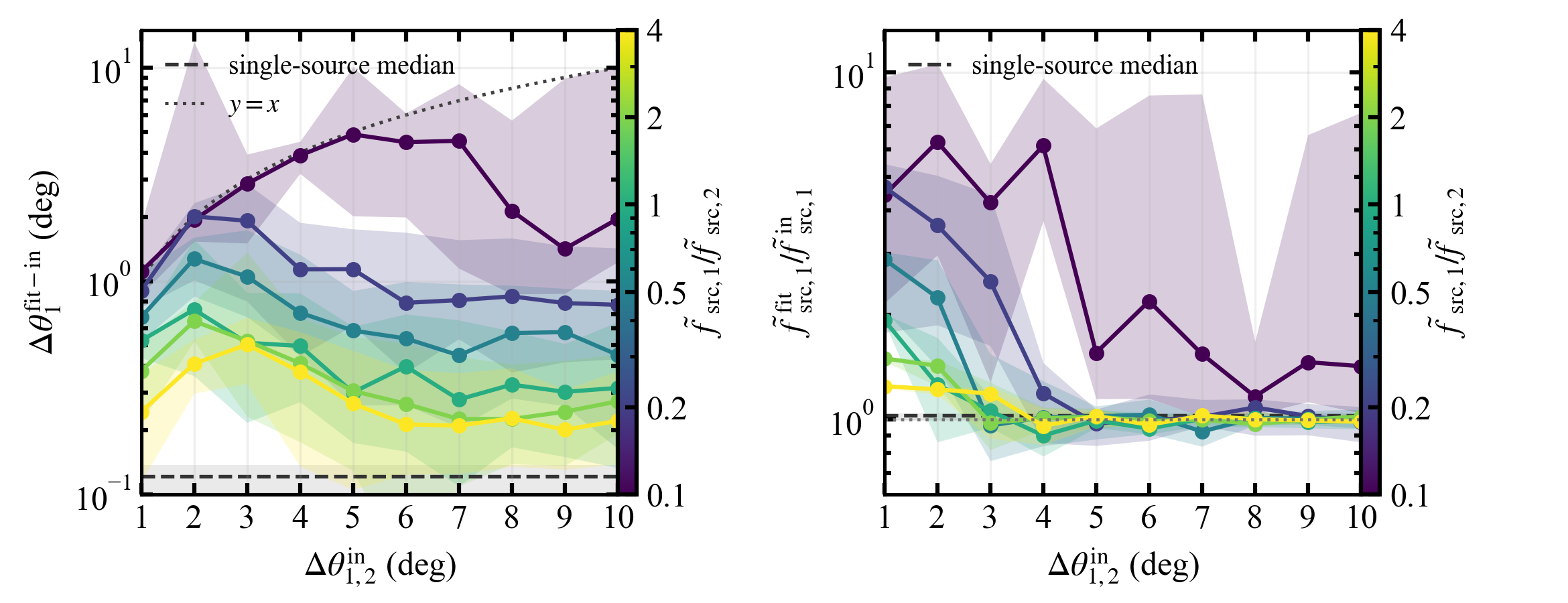}
      \caption{Performance of two-source fitting for recovering the position and flux of Source 1. The input angular separation between the two sources is denoted as $\Delta\theta_{1,2}^{\rm in}$, and different colors indicate the input flux ratio $\tilde f_{\rm src,1}/\tilde f_{\rm src,2}$. \textit{Left}: angular separation between the fitted and input positions of Source 1, $\Delta\theta_{1}^{\rm fit-in}$. The dotted diagonal line marks $\Delta\theta_{1}^{\rm fit-in}=\Delta\theta_{1,2}^{\rm in}$. \textit{Right}: recovered Source-1 flux normalized by the input value, $\tilde f_{\rm src,1}^{\rm fit}/\tilde f_{\rm src,1}^{\rm in}$. Solid curves show the median over 10 Poisson realizations, while shaded regions show the central 68\% realization scatter. The dashed horizontal line and gray band indicate the corresponding single-source fitting result and its central 68\% realization scatter.}
      \label{fig:twosrc}
  \end{figure}

A more complicated situation arises when multiple sources lie within the DIXE field of view. In this case, the localization precision can be degraded. Figure~\ref{fig:twosrc} shows the result of fitting two sources simultaneously, with the flux of Source 1 fixed at $\tilde f_{\rm src,1}=0.5~\phrate$. The flux ratio $\tilde f_{\rm src,1}/\tilde f_{\rm src,2}$ is varied from 0.1 to 4, and the input angular separation between the two sources, $\Delta\theta_{1,2}^{\rm in}$, is varied from 1 to 10 deg. As shown in Figure~\ref{fig:twosrc}, the localization precision of Source 1, $\Delta\theta_1^{\rm fit-in}$, is generally degraded in the presence of a second source. Despite a modest increase around $\Delta\theta_{1,2}^{\rm in}\sim2$ deg, the localization precision remains approximately stable with increasing source separation for $\tilde f_{\rm src,1}/\tilde f_{\rm src,2}\gtrsim0.2$. Results below the dotted line, $\Delta\theta_{1}^{\rm fit-in}=\Delta\theta_{1,2}^{\rm in}$, indicate that the localization error of Source 1 is smaller than the separation between the two sources, suggesting that the source position is reasonably well constrained. When the two sources are separated by less than about 3 deg and the target source is fainter than, or comparable to, the second source ($\tilde f_{\rm src,1}/\tilde f_{\rm src,2}\lesssim1$), the method may have difficulty distinguishing them.

\begin{figure}
  \centering
  \includegraphics[width=0.95\hsize]{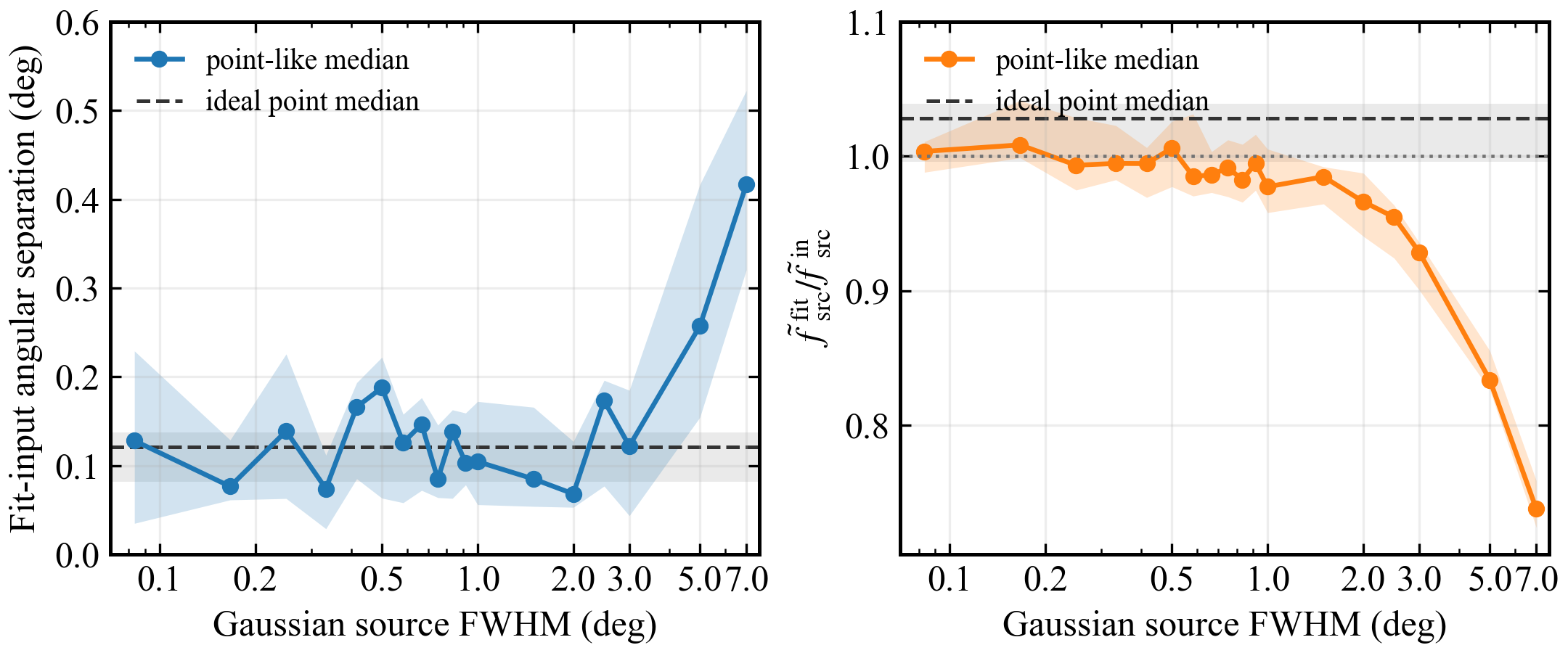}
  \caption{Effect of finite source extent on point-source fitting. Mock sources are generated as Gaussian profiles with different FWHM values and then fitted with the point-source response model. \textit{Left}: angular separation between the fitted and input source positions. \textit{Right}: recovered source flux normalized by the input value, $\tilde f_{\rm src}^{\rm fit}/\tilde f_{\rm src}^{\rm in}$. Solid curves show the median across the realizations, and the shaded regions indicate the central 68\% scatter across realizations. The dashed horizontal lines and gray bands show the corresponding ideal point-source results and their central 68\% scatter. The dotted line in the right panel marks unbiased flux recovery.}
  \label{fig:pointlike}
\end{figure}

So far we have assumed ideal point sources with no physical extent. This approximation is appropriate for the transient-localization application targeted here, because sources with intrinsic extents below the spatial sampling resolution of the method, here $10'$, are unresolved and are effectively absorbed into the flux assigned to the sampling pixel. Finite source extent only becomes relevant when the source size is larger than this sampling scale. Figure~\ref{fig:pointlike} shows a model-mismatch test in which Gaussian-like sources with FWHM ranging from 0.05 to 7 deg are fitted with the point-source model. In all cases, the integrated count rate is fixed at $0.5~\ctrate$. The median localization offset is $\sim0.1$ deg, comparable to the ideal point-source case, although the scatter becomes larger for sources with non-zero extent. The point-source fit begins to fail when ${\rm FWHM}\gtrsim3~{\rm deg}$. The recovered flux (Figure~\ref{fig:pointlike}, \textit{right}) becomes systematically lower than the true value when ${\rm FWHM}\gtrsim1~{\rm deg}$, because part of the source emission is sampled at off-axis angles where the collimator response has lower values (Figure~\ref{fig:cr_phase_space}, \textit{right}). Such highly extended and bright events are not expected for the point-like transient cases considered here, and an apparent extent of several degrees would generally rule out an interpretation as a point-like transient. From Figure~\ref{fig:pointlike} we conclude that finite source extent does not strongly bias the median localization precision before the fit starts to fail at large FWHM.



\subsection{Imaging capability for diffuse sources}
\label{imaging:diffuse}
\begin{figure}
    \centering
    \includegraphics[width=\hsize]{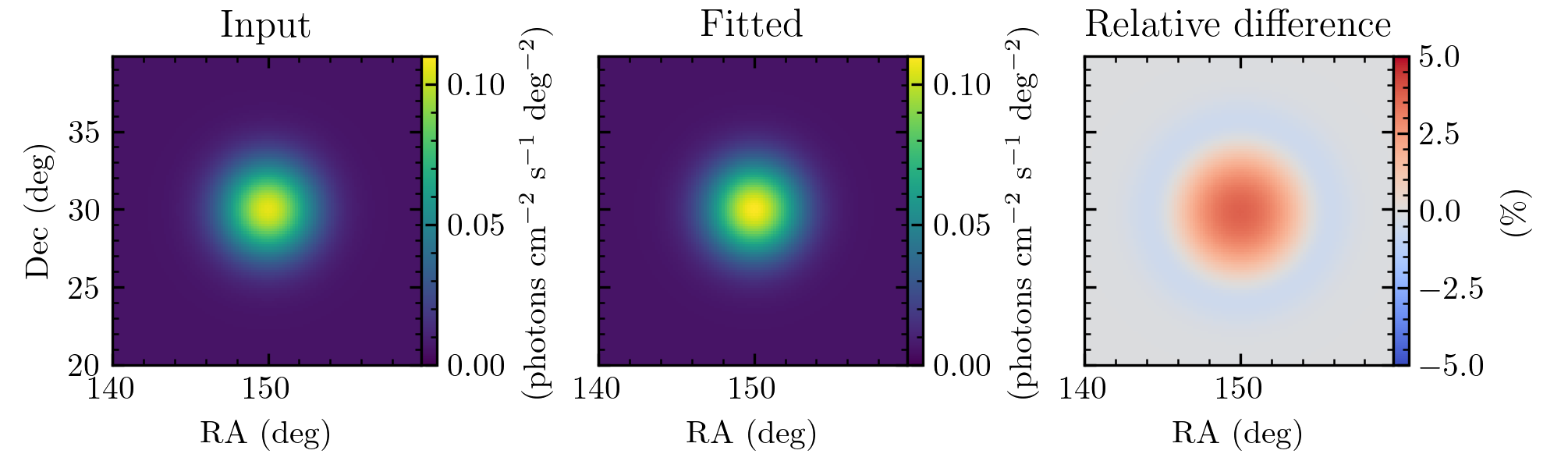}
    \caption{The input (\textit{left}) and best-fit (\textit{middle}) extended source profiles assume a Gaussian surface brightness distribution. The color coding represents surface brightness in units of $\phrate~\rm cm^{-2}~deg^{-2}$. The \textit{right} panel shows the relative difference between the best-fit and input profiles in percentage, with color indicating the value of \textit{(best-fit} $-$ \textit{input}) divided by \textit{input}.}
    \label{fig:fit_gauss}
\end{figure}

Diffuse sources have more complicated emission structures. Demodulation improves the imaging quality, illustrates the spatial distribution of the surface brightness, and provides insight into the underlying physical processes at different places. Simple extended sources where analytical models are sufficient to describe the surface brightness profile, the MCMC fitting is able to reconstruct the structure, similar to the point source case. An example is shown in Figure~\ref{fig:fit_gauss}. Assuming a Gaussian profile and a background level $\tilde F_{\rm bkg}=0.59~\ctrate$, the surface brightness distribution can be reconstructed with an $\lesssim5\%$ accuracy.

In more general cases, diffuse sources exhibit surface brightness structures that lack simple analytical descriptions. Features such as brightness discontinuities and changes in emission line ratios, as observed by eROSITA \citep{Predehl_2020_Nature,Zheng_2024_eRO_narrowband}, may indicate physical processes such as shock fronts or changes in plasma ionization states. Hot spots in the hard X-ray band may suggest regions where non-thermal X-ray emission dominates over thermal processes \citep{Ballet_2006_SNR_nonthermal,Mayer_2023_nonthermal_xray}. In these scenarios, rather than assuming a parametric form, we pixelate the image and directly fit the mean surface brightness in each pixel. The fitting approach for complex diffuse sources is illustrated in the inset of the \textit{left} panel of Figure~\ref{fig:psfsky_gauss_countrate}, where the model count rate is computed using Eq.~\eqref{eq:riemann_sum}.

\begin{figure}
    \centering
    \includegraphics[width=0.95\hsize, trim={1.5cm 0cm 0cm 0cm}]{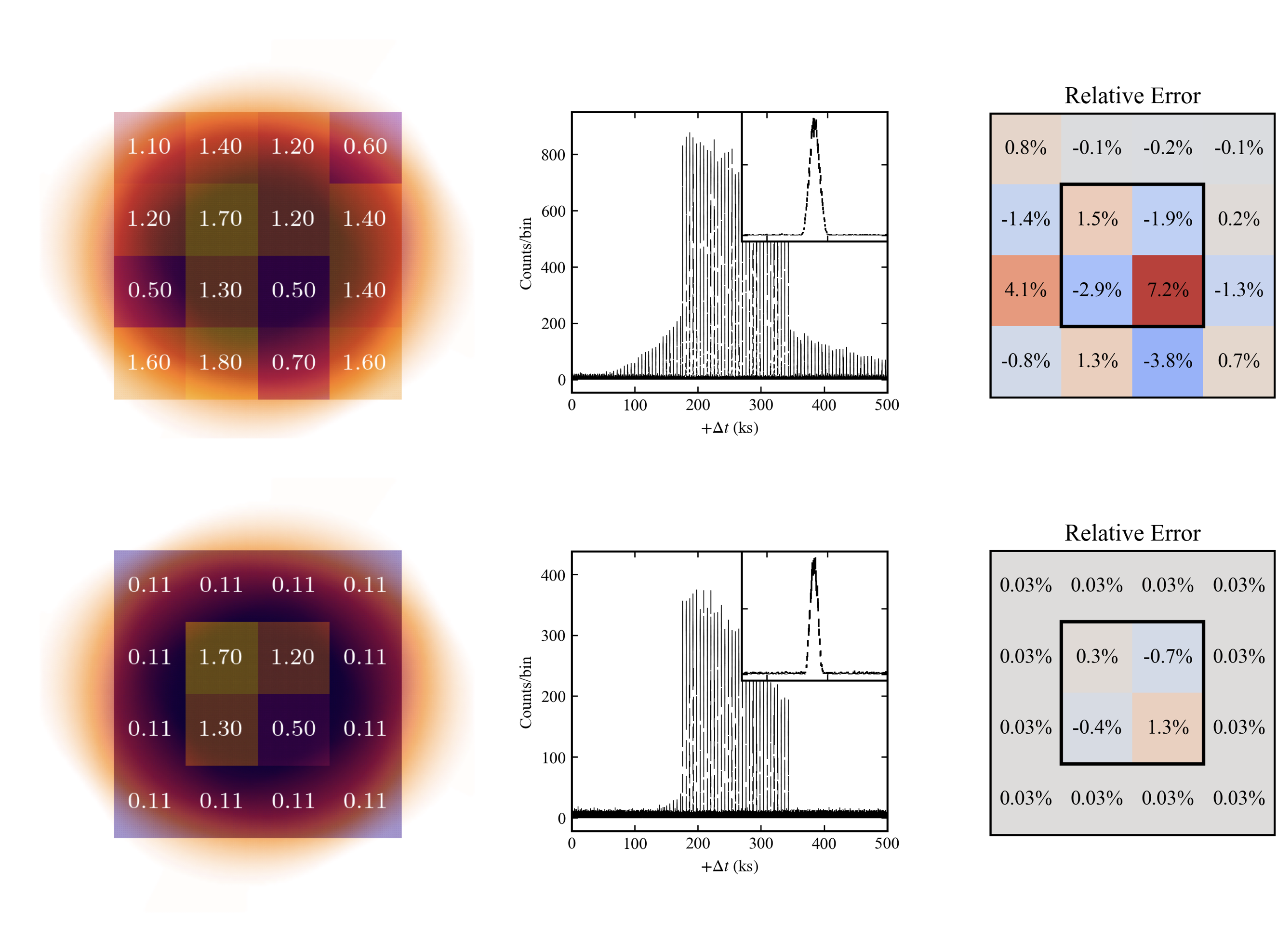}
    \caption{Input images (\textit{left}), mocked count-rate profiles (\textit{middle}), and relative errors between the best-fit and input images (\textit{right}). Upper panels: The left panel shows a $4\times4$ image with a pixel size of $3^\circ\times3^\circ$, overlaid with the collimator response pattern at the time when the DIXE field of view passes through the image center. The numbers indicate the mean surface brightness in units of $\phrate~\mathrm{cm}^{-2}~\mathrm{deg}^{-2}$. A uniform background level of $0.59~\rm \ctrate$ is assumed. The middle panel shows the corresponding mocked count-rate profile over a four-day interval; the inset displays a zoomed-in view of a single scan. The right panel shows the relative error between the best-fit and input images, with values given in percent. Lower panels: Same as the upper panels, but for a $2\times2$ image (with the same pixel size) embedded in a uniform background. The surface brightness of the $2\times2$ image is identical to that of the central $2\times2$ pixels in the upper-left panel. Black boxes in the right panels highlight the corresponding pixels to guide the eye.}
    \label{fig:fit_image}
\end{figure}

Similar to the point-source case, the MCMC fitting uncertainties depend on the exposure time. For method verification, we use 4-day mock data. 
Figure~\ref{fig:fit_image}, \textit{upper} panel shows the fitting results for a $4\times4$ pixel image with each pixel spanning $3^\circ\times3^\circ$. A uniform background level of $0.59~\ctrate$ is assumed. As shown in Figure~\ref{fig:fit_image}, we achieve a relative difference between the input and fitted images of less than $8\%$, given the assumed surface brightness level. 

It is important to note that for diffuse sources, the assumption of a uniform background is simplistic. To investigate the impact of background variation, we also perform a fit on a $2\times2$ image using the same surface brightness values as the central $2\times2$ region of the $4\times4$ image, shown in the \textit{lower left} panel of Figure~\ref{fig:fit_image}. By comparing the relative errors in the central $2\times2$ pixels (Figure~\ref{fig:fit_image}, \textit{right} panel), we find that three out of four pixels yield lower relative errors when assuming a uniform, low background. However, the results also indicate that when treating the central $2\times2$ region as the target and surrounding pixels as part of a random background field, the fitting accuracy remains within $5\%$, demonstrating the method’s feasibility. In practical applications, one could iteratively fit an initial region pixel by pixel and expand outward from it.

As the image dimension increases, the MCMC fitting time scales almost linearly with the number of pixels. However, reducing pixel size dramatically increases computational time. It is also worth noting that the surface brightness level used in Figure~\ref{fig:fit_image} is relatively high; therefore, this method is most applicable to bright diffuse sources where a detailed spatial distribution of photons is scientifically valuable, such as in the Vela–Puppis A supernova remnant.

\section{Summary}
We optimized the DIXE survey strategy by evaluating various pointing inclination angles and solar avoidance strategies. A pointing inclination angle of $0^\circ$ yields the maximum sky coverage, approximately 72.5\%. At this configuration, DIXE covers most of its main science targets and important calibration sources.

To mitigate solar contamination, the angular separation between the DIXE FoV and the Sun must exceed $20^\circ$. We define two Sun avoidance strategies: FD, which minimizes mechanical risk by reducing the frequency of Sun shield operations, and FI, which maximizes observational efficiency. The typical one-year exposure time under FD and FI is 26 ks and 68 ks, respectively. The mission operation will start with the FD strategy and switch to the FI strategy progressively. Under the FI strategy, the survey period is $\sim$70 days. 

Incorporating the optimized survey strategy and the DIXE collimator response, we apply a demodulation method based on Markov Chain Monte Carlo (MCMC) sampling to enhance DIXE’s imaging capability. For point sources such as unanticipated transient events, the method yields a localization accuracy of $\sim1^\circ$ for sources with observed fluxes above $0.1~\rm photons~s^{-1}~cm^{-2}$ and background levels below $0.59\rm ~counts~s^{-1}$ in $2-8~\rm keV$. For two-point-source cases, source confusion becomes important when two sources are separated by less than about $3^\circ$, especially when the target source is fainter than, or comparable to, the second source. Source extents below the sampling scale of the method do not affect the point-source fit, while larger extents are expected to be rare for transient applications. Even in the model-mismatch tests, the median localization remains comparable to the ideal point-source case until the fitted source becomes very extended, with clear failure appearing only for ${\rm FWHM}\gtrsim3^\circ$; such an extent would generally exclude a point-like transient origin.

For extended sources, improved spatial resolution can reveal internal structures and shed light on underlying physical mechanisms. When modeled with a Gaussian surface brightness profile, the method achieves surface brightness constraints within a $5\%$ relative error for sources with a central brightness of $1~\rm photons~s^{-1}~cm^{-2}~deg^{-2}$ and background of $0.59~\rm counts~s^{-1}$ in the $2-8~\rm keV$ band.
For more realistic cases without analytical descriptions, we perform pixel-based fitting of the surface brightness distribution. A test on a $4\times4$ pixel image (each $3^\circ \times 3^\circ$) with $1~\rm counts~s^{-1}$ background yields a relative deviation below 8\% using four-day data. The uncertainties can be improved with a longer exposure. While computationally intensive, this method is suitable for bright sources where detailed spatial information is desired.

\label{sec:summary}

\section*{Acknowledgment}
This work was supported in part by the National Natural Science Foundation of China (Grant No. 12220101004, 11821303) and the Ministry of Science and Technology of China (Grant No. 2018YFA0404502). J. M. acknowledges support from the Tsinghua Dushi Program 53121200125. We acknowledge support from the ISSI-BJ forum: "Surveying the Hot Baryons Within the Milky Way". We acknowledge the Tsinghua Astrophysics High-Performance-Computing (TAHPC) platform for providing computational and data storage resources. \textbf{}

\bibliographystyle{raa}
\bibliography{ref}

\begin{appendices}


\section{Background levels}
\label{sec:bkg}
The point sources generally have hard spectra ($>2~\rm keV$). Therefore, we calculate the background in the range of $2-8~\rm keV$. For the background component, cosmic X-ray background (CXB) and non-X-ray background \citep[NXB;][]{Tian_2026} from the secondary particles generated by cosmic rays are considered.

\begin{figure}
    \centering
    \includegraphics[width=0.5\hsize]{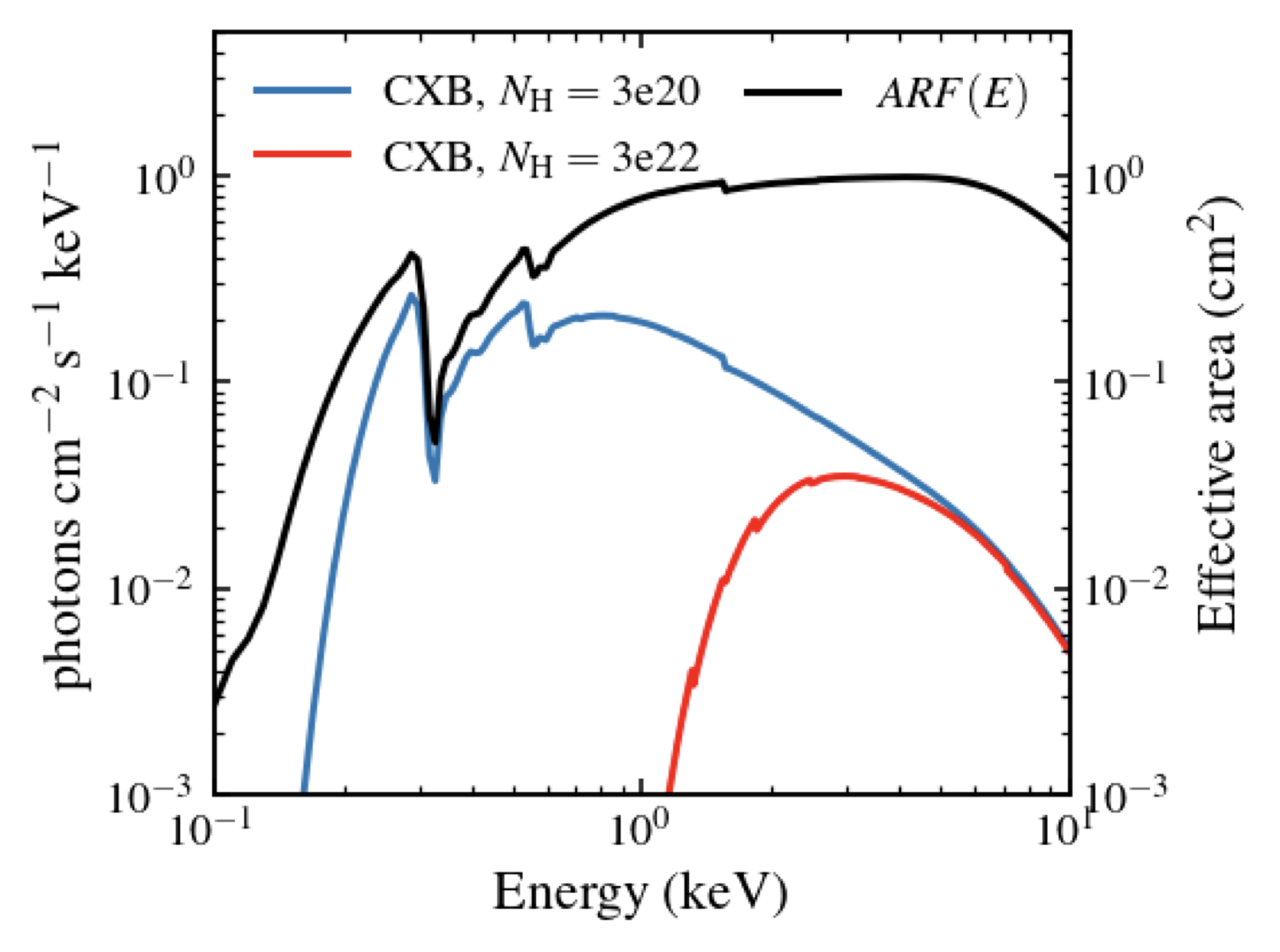}
    \caption{Effective area convolved CXB spectra assuming an absorbing neutral hydrogen column density $N_{\rm H}=3\times10^{20}~\rm cm^{-2}$ (blue) and $N_{\rm H}=3\times10^{22}~\rm cm^{-2}$ (red). The solid black line shows the DIXE effective area $A_{\rm eff}(E)$ of DIXE filters in $0.1-10~\rm keV$.}
    \label{fig:cxb}
\end{figure}

The photon index of CXB is $\Gamma=1.41\pm0.06$ and a surface brightness in energies $2-8~\rm keV$ of $(2.24\pm0.16)~\rm erg~cm^{-2}~s^{-1}~deg^{-2}$ \citep{DeLuca_2004_AA_CXB}. Therefore, DIXE with a FoV of $100~\rm deg^2$, receives a CXB flux of $\simeq2.24\times10^{-9}~\rm erg~\rm cm^{-2}~s^{-1}$. Adopting $\Gamma=1.41$, it corresponds to a normalization of $0.27~\rm photons\rm ~cm^{-2}~s^{-1}~\rm keV^{-1}$ at 1~keV. Without loss of generality, we assume a neutral column density of $N_{\rm H}=3\times10^{22}~\rm cm^{-2}$ and $3\times10^{20}~\rm cm^{-2}$ as an estimation for the sources in and out of the Galactic plane. Attenuation due to VIS/UV/IR blocking filters is also considered. Figure \ref{fig:cxb} shows the CXB spectra ($N_{\rm H}$ absorbed) before and after passing through the filters. The integration of the filter-attenuated CXB spectra in the range $2-8~\rm keV$ gives $0.21$ and $0.15 ~\rm photons~cm^{-2}~\rm s^{-1}$ for the assumed low and high $N_{\rm H}$, respectively. The area of the TES detector array is $1~\rm cm^2$. Thus the CXB count rate received by DIXE is $F_{\rm bkg,CXB}\simeq0.21~\rm photons~s^{-1}$ and $0.15~\rm photons~s^{-1}$. We adopt a typical NXB level of $0.38~\ctrate$ in the $2-8~\rm keV$ band from Geant4 simulation \citep{Tian_2026}. The total background count rates, including the NXB and CXB, are shown in Table~\ref{tab:bkg}.


\section{MCMC fitting}
\label{sec:mcmc}
\begin{figure}
    \centering
    \includegraphics[width=0.9\hsize]{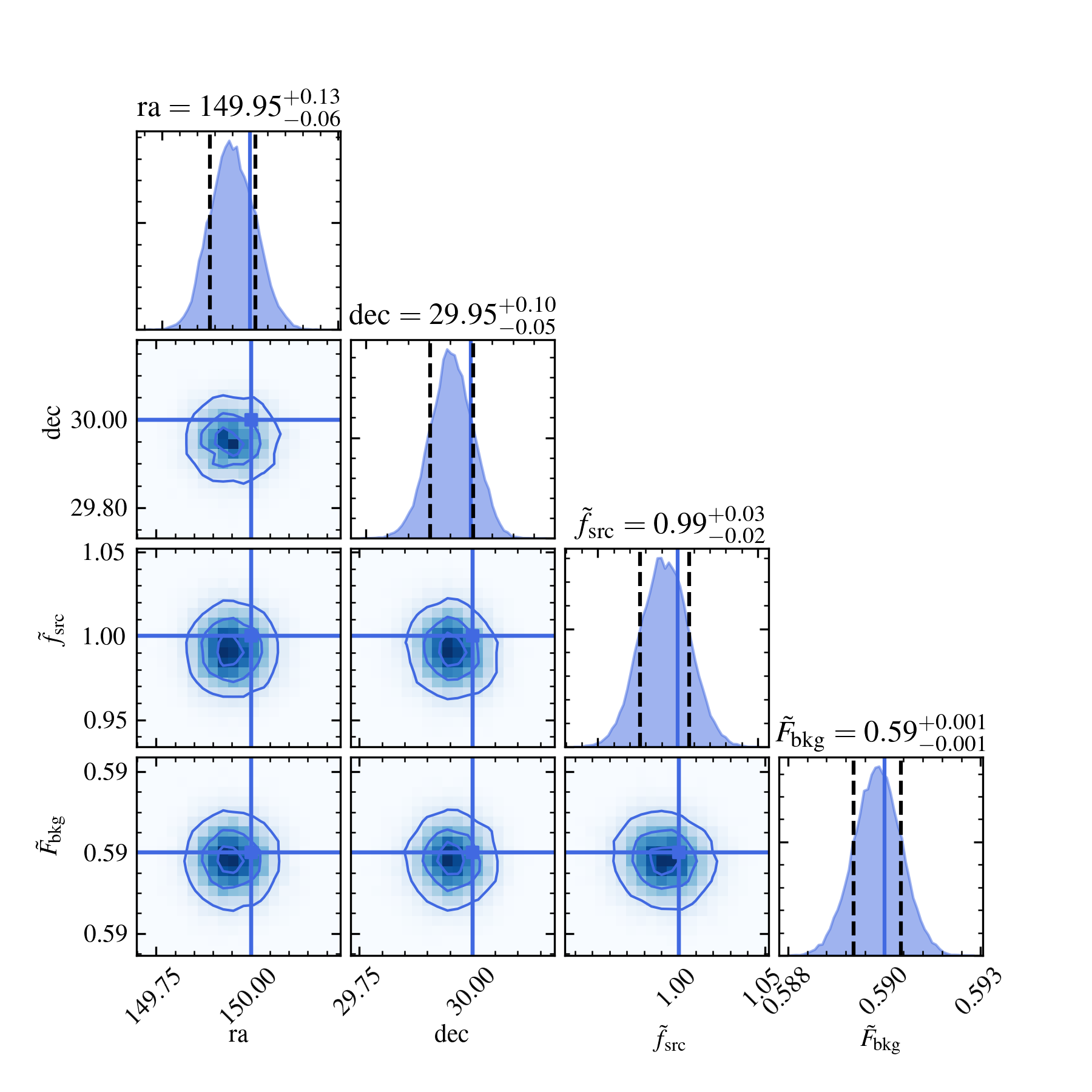}
    \caption{The corner plot of the MCMC fitting results with $\tilde f_{\rm src}=1~\phrate$ and $\tilde F_{\rm bkg}=0.59~\ctrate$. The solid blue lines show the true value and the dashed black lines enclose the fitted values within 1$\sigma$ confidence level. The best-fit values with 1$\sigma$ error are written.}
    \label{fig:ps_mcmc_corner}
\end{figure}

\begin{figure}
    \centering
    \includegraphics[width=\hsize]{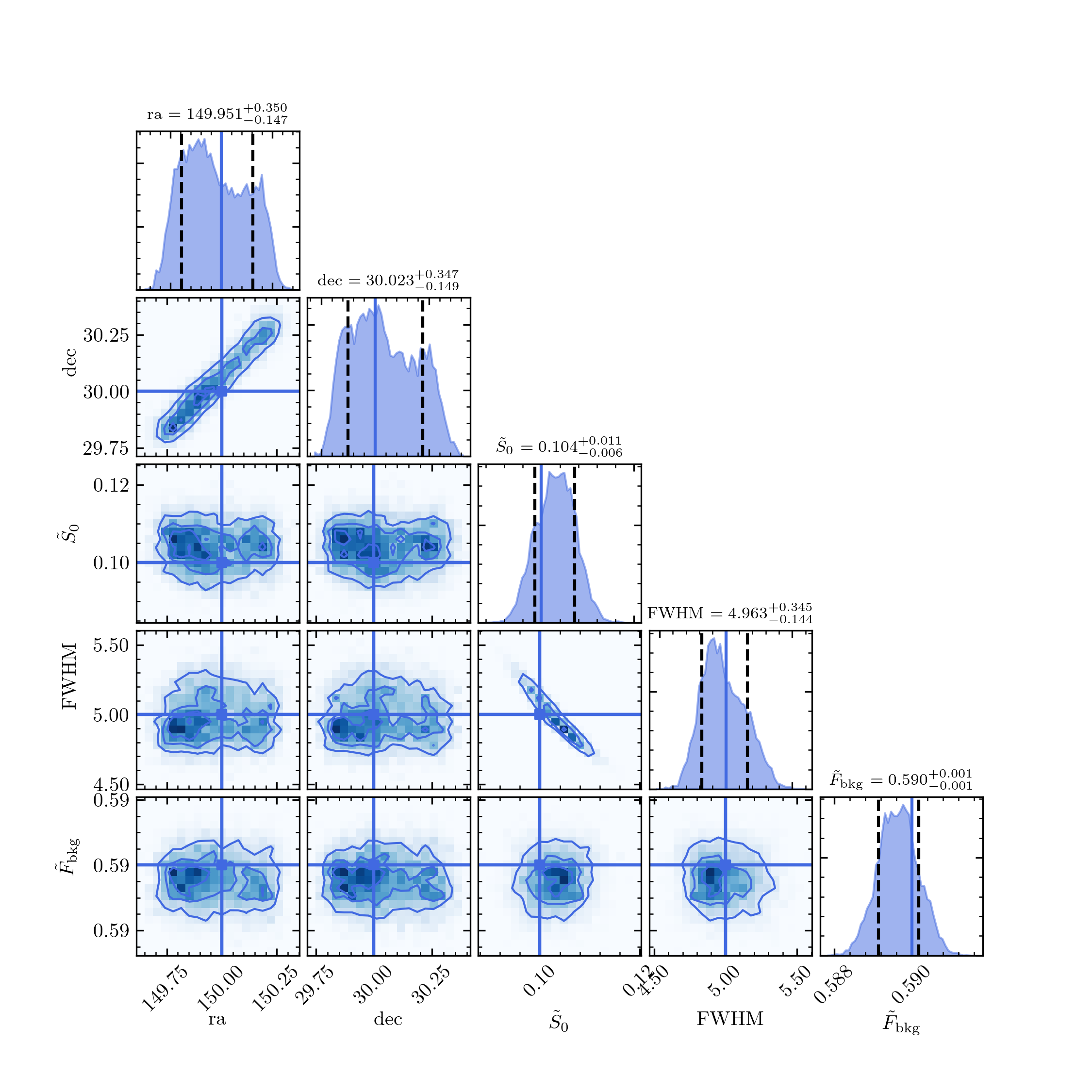}
    \caption{Similar to Figure~\ref{fig:ps_mcmc_corner} and shows the fitting results for an extended source with a Gaussian profile.}
    \label{fig:gauss_mcmc_corner}
\end{figure}

We perform the MCMC fitting for the three different kinds of science targets described in Section~\ref{sec:science_capability} using the python package \textit{emcee} \cite{Foreman-Mackey_2013_emcee}. Figure~\ref{fig:ps_mcmc_corner} shows a typical corner plot of the MCMC fitting result for point-like sources described in Section~\ref{imaging:ps}. Figure~\ref{fig:gauss_mcmc_corner} shows the results for an extended source with a Gaussian profile. The central surface brightness $S_0=0.1~\rm \phrate~cm^{-2}~deg^{-2}$ and the FWHM is $5^\circ$.

\end{appendices}



\end{document}